\documentclass[acmsmall,screen]{acmart}

\setcopyright{rightsretained}
\acmPrice{}
\acmDOI{10.1145/3276491}
\acmYear{2018}
\copyrightyear{2018}
\acmJournal{PACMPL}
\acmVolume{2}
\acmNumber{OOPSLA}
\acmArticle{121}
\acmMonth{11}

\usepackage{soul}
\usepackage{booktabs}
\usepackage{subcaption}
\usepackage{xspace}
\usepackage[normalem]{ulem}
\usepackage{listings}%
\usepackage{xcolor}
\usepackage[font=small,labelfont=bf,aboveskip=2pt,belowskip=0pt]{caption}
\usepackage{parcolumns}
\usepackage{algorithm}
\usepackage{algorithmicx}
\usepackage[noend]{algpseudocode}
\usepackage{enumitem}
\newcommand{\defn}[1]{\textbf{#1}}
\usepackage{bm}

\newcommand{\myparagraph}[1]{\vspace{0.5pt} \noindent {\bf #1.}}
\newcommand\punt[1]{}
\newcommand{\performanceInconsistency}{performance inconsistency\xspace}

\newcommand{\graphit}{GraphIt\xspace}

\newcommand{\julian}[1]{{\color{blue} {\bf Julian:} #1}}

\newcommand{\yunming}[1]{{\color{red} {\bf Yunming:} #1}}
\newcommand{\shoaib}[1]{{\color{green} {\bf Shoaib:} #1}}

\newcommand{\updated}[1]{#1}

\newcommand{\apply}{{\small\texttt{apply}\xspace}}
\newcommand{\applyModified}{{\small\texttt{applyModified}\xspace}}
\newcommand{\too}{{\small\texttt{to}\xspace}}
\newcommand{\from}{{\small\texttt{from}\xspace}}
\newcommand{\srcFilter}{{\small\texttt{srcFilter}\xspace}}
\newcommand{\dstFilter}{{\small\texttt{dstFilter}\xspace}}
\newcommand{\scheduleKeyword}{{\small\texttt{schedule}\xspace}}

\newcommand{\fuseApplyFunctions}{{\small\texttt{fuseApplyFunctions}\xspace}}
\newcommand{\numExp}{32\xspace}
\newcommand{\numFastExp}{24\xspace}

\newcommand{\work}{work-efficiency\xspace}
\newcommand{\Work}{Work-efficiency\xspace}

\newcommand{\Fig}{Fig.}

\newcommand{\SSG}{SSG\textunderscore ID\xspace}
\newcommand{\BSG}{BSG\textunderscore ID\xspace}

\newcommand{\outerIter}{OuterIter\xspace}
\newcommand{\innerIter}{InnerIter\xspace}

\newcommand{\SparsePush}{SparsePush\xspace}
\newcommand{\DensePush}{DensePush\xspace}
\newcommand{\DensePull}{DensePull\xspace}

\newcommand{\stt}[1]{{\small \texttt{#1}}}
\newcommand{\ftt}[1]{{\footnotesize \texttt{#1}}}

\newenvironment{denseitemize}{
\begin{itemize} [topsep=2pt, partopsep=0pt, leftmargin=1.5em]
 \setlength{\topsep}{0pt}
 \setlength{\itemsep}{2pt}
 \setlength{\parskip}{0pt}
 \setlength{\parsep}{0pt}
}{\end{itemize}}

\let\origthelstnumber\thelstnumber
\makeatletter
\newcommand*\Suppressnumber{%
  \lst@AddToHook{OnNewLine}{%
    \let\thelstnumber\relax%
     \advance\c@lstnumber-\@ne\relax%
    }%
}

\newcommand*\Reactivatenumber[1]{%
  \setcounter{lstnumber}{\numexpr#1-1\relax}
  \lst@AddToHook{OnNewLine}{%
   \let\thelstnumber\origthelstnumber%
   \refstepcounter{lstnumber}
  }%
}
\makeatother

\usepackage{microtype}
\usepackage{subcaption,wrapfig}
\definecolor{gray}{gray}{0.5}
\definecolor{key}{rgb}{0,0.5,0}

\def\OPTL{\textrm{$[$}}
\def\OPTR{\textrm{$]$}}

\definecolor{lightbackground}{rgb}{.98,.98,.97}
\definecolor{darkgray}{rgb}{.3,.3,.3}
\definecolor{darkred}{rgb}{.6,0,0}
\definecolor{darkgreen}{rgb}{0,.6,0}
\definecolor{darkblue}{rgb}{0,0,.6}

\lstdefinelanguage{graphit}{%
  keywords={[1]},
  keywords={[2]func,end,element,const,var, field,for,%
    vertexset,edgeset,vector,%
    void,char,short,long,int,float,double,boolean,size_t},
  keywords={[3]filter, from, to, srcFilter, dstFilter, apply, applyModified},
  keywords={[4],schedule, s1, l1, l2, l3 %
    },
  literate={[OPT[}{{\OPTL}}1 {]OPT]}{{\OPTR}}1,
  string=[b]",
  comment=[l]//,
  morecomment=[s]{/*}{*/},
  mathescape=true,
  flexiblecolumns=true,
  tabsize=2,
  captionpos=b,
  frame=single,
  framerule=0pt,
  aboveskip=2pt,
  belowskip=1pt,
  framesep=1pt,
  basicstyle=\scriptsize\ttfamily,
  keywordstyle={[1]\color{darkred}},
  keywordstyle={[2]\color{blue}},
  keywordstyle={[3]\color{black}\bfseries},
  keywordstyle={[4]\color{darkred}\bfseries},
  numbers=left,
  stepnumber=1,
  numbersep=3pt,
  numberstyle=\tiny,
  emphstyle=\slshape,
  commentstyle=\color{darkgray},
  stringstyle=\color{darkgreen},
  xleftmargin=4.0ex,
}

\begin{document}

\title{GraphIt: A High-Performance Graph DSL}


\author{Yunming Zhang}
\affiliation{
  \position{}
  \department{}             
  \institution{MIT CSAIL}           
  \country{USA}                   
}
\email{yunming@mit.edu}          

\author{Mengjiao Yang}
\affiliation{
  \position{}
  \department{}             
  \institution{MIT CSAIL}           
  \country{USA}                   
}
\email{mengjiao@mit.edu}          

\author{Riyadh Baghdadi}
\affiliation{
  \position{}
  \department{}             
  \institution{MIT CSAIL}           
  \country{USA}                   
}
\email{baghdadi@mit.edu}          

\author{Shoaib Kamil}
\affiliation{
  \position{}
  \department{}             
  \institution{Adobe Research}           
  \country{USA}                   
}
\email{kamil@adobe.com}          

\author{Julian Shun}
\affiliation{
  \position{}
  \department{}             
  \institution{MIT CSAIL}           
  \country{USA}                   
}
\email{jshun@mit.edu}          

\author{Saman Amarasinghe}
\affiliation{
  \position{}
  \department{}             
  \institution{MIT CSAIL}           
  \country{USA}                   
}
\email{saman@csail.mit.edu}          

\begin{abstract}

The performance bottlenecks of graph applications
depend not only on the
algorithm and the underlying hardware, but also on
the size and structure of the input graph.
As a result, programmers must try different combinations of a large
set of techniques, which
make tradeoffs among locality, \work, and parallelism, to develop the best implementation
for a specific algorithm and type of graph.
Existing graph frameworks and domain specific languages (DSLs) lack flexibility, supporting
only a limited set of optimizations.

This paper introduces \textbf{\graphit}, a new DSL
for graph computations that generates fast implementations for algorithms
with different performance characteristics running on graphs
with different sizes and structures.
\graphit separates what is computed
({algorithm}) from how it is computed ({schedule}).
Programmers specify the algorithm using an \emph{algorithm
language}, and performance optimizations are specified using a
separate \emph{scheduling language}.
The algorithm language simplifies expressing the algorithms,
while exposing opportunities for optimizations.
We formulate graph optimizations, including edge traversal direction, data layout, parallelization, cache, NUMA, and kernel fusion optimizations, as tradeoffs among
locality, parallelism, and \work.
The scheduling language enables programmers
to easily search through this complicated
tradeoff space by composing together a large
set of edge traversal, vertex data layout, and program structure optimizations.
 The separation 
 of algorithm and schedule also enables us to build
 an autotuner on top of \graphit to automatically find 
 high-performance schedules. 
The compiler uses a new scheduling representation, the
\textit{graph iteration space}, to model,
compose, and ensure the validity of the large number of
 optimizations.
We evaluate \graphit's performance with seven algorithms on
graphs with different structures and sizes.
\graphit outperforms the next fastest of six state-of-the-art shared-memory
frameworks (Ligra, Green-Marl, GraphMat, Galois, Gemini, and Grazelle)
on \numFastExp out of \numExp experiments by up to 4.8$\times$, and is
 never more than 43\% slower than the fastest framework on the other experiments. 
 \graphit also reduces the lines of code by up to an order of magnitude compared
 to the next fastest framework.

\punt{
The separation of the algorithm from the schedule enables programmers to
try different combinations of optimizations, using a separate scheduling language,
without changing the high-level algorithm.}

\punt {

They can also find fast schedules by auto-tuning the scheduling language or simply exploring the full space of schedules.
}

\end{abstract}



\maketitle

\renewcommand{\shortauthors}{Y. Zhang, M. Yang, R. Baghdadi, S. Kamil, J. Shun, S. Amarasinghe}

\section{Introduction}
\label{sec:intro}

In recent years, large graphs with billions of vertices and trillions
of edges have emerged in many domains, such as social network
analytics, machine learning, and biology.  Extracting information from
these graphs often involves running algorithms for identifying
important vertices, finding connections among vertices, and detecting
communities of interest.  Speeding up these algorithms can enhance the
efficiency of data analytics applications and improve the quality of
web services~\cite{Sharma2016,Zhisong2017,Eksombatchai2018}.

\punt{ As a result, there has been significant interest in developing
  high-performance graph processing frameworks.\shoaib{Too much
    passive voice.  Is the important thing the renewed interest, or
    the need for high performance?}

  \shoaib{throughout the intro, unclear why we should care about
    "performance characteristics".  perhaps what you mean to say is
    that the performance bottlenecks are different for each algorithm,
    so it requires doing different optimizations?  "performance
    characteristics" is too vague a term, and furthermore the
    implication is that the algorithm has inherent "characteristics"
    separate from the specific machine they're running on.  But these
    measures will be different for different machines} It is
  challenging to get high performance for different graph algorithms.
  The performance characteristics of graph algorithms, such as cache
  miss rate, memory level parallelism and branch miss-predication
  rates are very different, and depend on the size and structure of
  the graph~\cite{Beamer15IISWC}.  }

It is difficult to implement high-performance graph algorithms. The
performance bottlenecks of these algorithms depend not only on the
algorithm and the underlying hardware, but also on the size and
structure of the graph~\cite{Beamer15IISWC}. As a result, different
algorithms running on the same machine, or even the same algorithm
running with different types of graphs on the same machine, can
exhibit different performance bottlenecks.  For example, some algorithms, such as
PageRank, spend most of the time working on the entire graph, while
traversal algorithms, such as Breadth-First Search (BFS), work on a
smaller subgraph at a time. In addition, running the same algorithm on a social
network with small diameter and power-law degree distribution exposes
different performance bottlenecks compared to running on a road
network with large diameter and uniform degree distribution.

Graph optimizations make tradeoffs among locality, parallelism and
\work~\cite{Kiriansky2016, Yunming2017, Beamer2017, shun13ppopp-ligra,
  Beamer-2012} to improve performance.  Since many graph algorithms
have very different performance bottlenecks, and optimizations make
different tradeoffs, one optimization can significantly boost the
performance for some graph algorithms on certain types of graphs,
while hurting the performance of other algorithms or the same
algorithm running on other types of graphs~\cite{jasmina2017}.
Programmers must iterate over multiple implementations of the same
algorithm to identify the best combination of optimizations for a
specific algorithm and input data.

\punt{ While the effects of graph optimization techniques have been
  known to expert programmers, one of the contributions of this paper
  is formulating graph optimizations as tradeoffs among locality,
  parallelism, and \work (Section~\ref{sec:tradeoff}).  }

\punt{ Combining together optimizations using conditionals also adds
  significant runtime overhead.}

Existing graph frameworks perform well for a subset of algorithms for
specific types of input, but have suboptimal performance on algorithms
with different bottlenecks and graphs of different sizes and
structures~\cite{Beamer15IISWC, Satish14Sigmod}.  Developers writing
performance-critical applications cannot afford such a level of
slowdown.  This performance inconsistency exists because each
framework was designed to support only a limited set of optimization
techniques, and does not allow for easy exploration of the large space
of optimizations.  It is infeasible to write hand-optimized code for
every combination of algorithm and input type.  A compiler approach
that generates efficient implementations from high-level specifications
is therefore a good fit.  However, existing graph domain specific
languages (DSLs)~\cite{Hong12asplos,socialite13ICDE,Aberger2016} do
not support composition of optimizations or expose comprehensive
performance tuning capabilities to programmers.

\punt{
  Tackling this \performanceInconsistency poses two major challenges
  for the design of the frameworks.  First, the framework would need a
  programming model that exposes optimization opportunities, while
  providing an interface to specify and combine a large number of
  optimizations.  Additionally, the framework requires a uniform
  scheduling representation that can represent, compose, and reason
  about the validity of different combinations of optimizations.  }

\punt{ single and multi-socket \yunming{4-socket numbers?}}  We
introduce \graphit,\footnote{The \graphit compiler is available under the MIT license at 
\url{http://graphit-lang.org/}} 
a new graph DSL that produces efficient
implementations with performance competitive with or faster than
state-of-the-art frameworks for a diverse set of algorithms running on
graphs with different sizes and structures. \graphit achieves good
performance by enabling programmers to easily find the best
combination of optimizations for their specific algorithm and input
graph. In this paper, we focus on providing mechanisms that make it
possible to easily and productively explore the space of
optimizations.

\graphit separates algorithm specifications from the choice of
performance optimizations.  Programmers specify the algorithm using an
\emph{algorithm language} based on high-level operators on sets of
vertices and edges. Performance optimizations are specified using a
separate \emph{scheduling language}.  The algorithm language
simplifies expressing algorithms and exposes opportunities for
optimizations by separating edge processing logic from edge traversal,
edge filtering, vertex deduplication, and synchronization logic.  We
formulate graph optimizations, including edge traversal direction,
data layout, parallelization, cache, NUMA, and kernel fusion optimizations, as
tradeoffs among locality, parallelism and \work.  The scheduling
language enables programmers to easily search through the complicated
tradeoff space by composing a large set of edge traversal, vertex
data layout, and program structure optimizations.

\graphit introduces novel scheduling representations for edge traversal,
vertex data layout, and program structure optimizations. 
Inspired by iteration space theory
for dense loops~\cite{wolf1991loop, padua1986advanced}, we introduce
an abstract \emph{graph iteration space} model to represent, compose,
and ensure the validity of edge traversal optimizations.  We encode
the graph iteration space in the compiler's intermediate
representation to guide program analyses and code
generation.

\updated{
The separation of algorithm and schedule also
 enables GraphIt to search for high-performance schedules automatically. 
 The large scheduling space and long running time of the applications 
 make it costly to do an exhaustive search. 
 We show that it is possible to discover schedules with good performance in 
 much less time using autotuning.
 Programmers familiar with graph optimizations 
 can leverage their expertise to tune 
 performance using the scheduling language directly.
} 

\punt{
In this paper, we focus on high-performance graph computing on
multicore shared-memory systems.  Increasingly, many graph frameworks
target shared-memory multicore
machines~\cite{shun13ppopp-ligra,zhang15ppopp-numa-polymer,sundaram15vldb-graphmat,nguyen13sosp-galois,Grossman2018,Sun2017}
because they have the smallest communication overheads, and memories
have grown to the point where many graphs can fit in a single
server. Shared-memory graph systems have also been shown to be more
efficient than distributed graph processing
systems~\cite{mcsherry15hotos}.
}

\begin{figure}[t]
  \centering 
  \includegraphics[width=\textwidth]{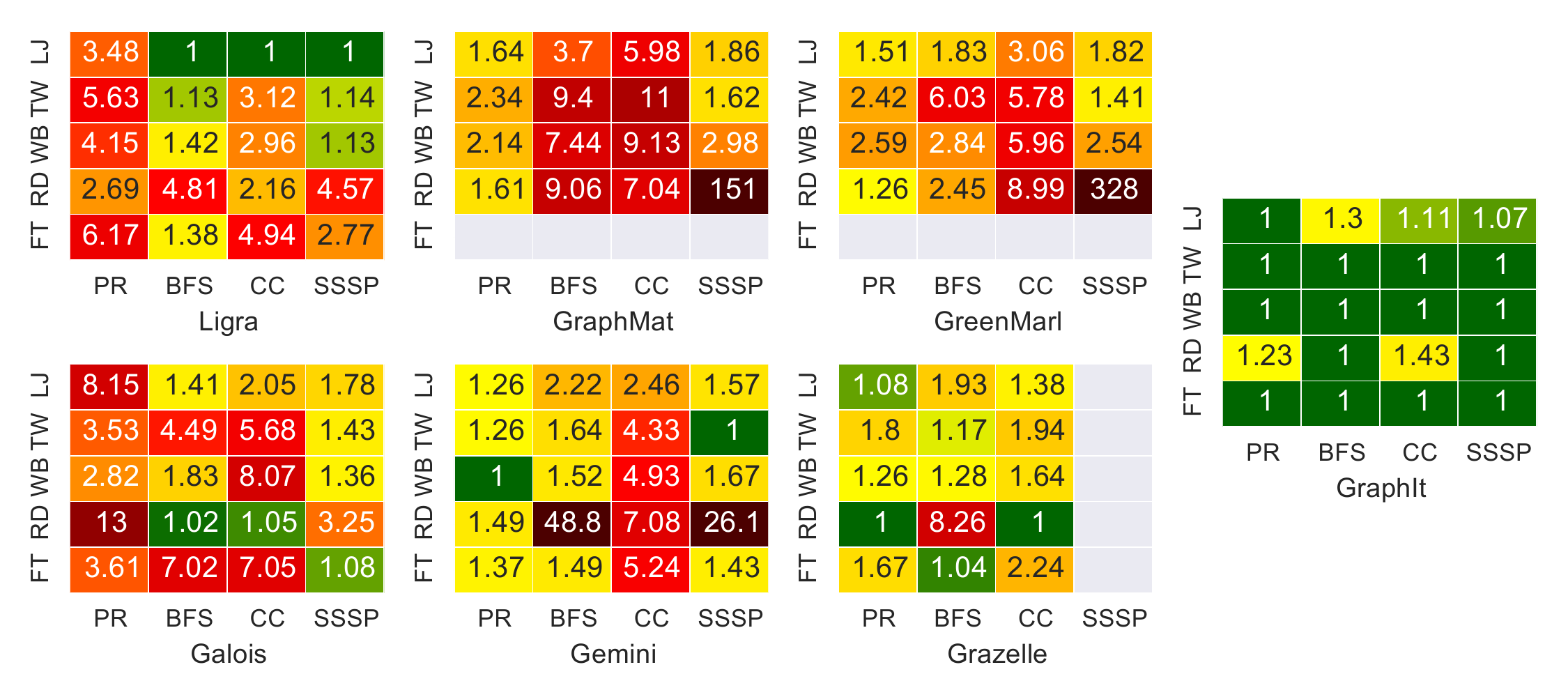}
  \caption{A heat map of slowdowns of various 
    frameworks compared to the fastest of all frameworks for PageRank
    (PR), Breadth-First Search (BFS), Connected Components (CC) using
    label propagation, and Single Source Shortest Paths (SSSP) using
    Bellman-Ford, on five graphs with varying sizes and structures
    (LiveJournal (LJ), Twitter (TW), WebGraph (WB), USAroad (RD),
    and Friendster (FT)). Lower numbers (green) are better, with one
    being the fastest for the specific algorithm running on the
    specific graph. Gray means that either an algorithm or a graph is not
    supported by the framework. We try to use the same algorithms across
    different frameworks. For Galois, we used the asynchronous algorithm 
    for BFS, and the Ligra algorithm for SSSP.}
  \label{fig:intro_comparisons}
  \vspace{3pt}
\end{figure}

We perform a comprehensive analysis of \graphit's performance using
six state-of-the-art shared-memory frameworks
(Ligra~\cite{shun13ppopp-ligra},
GraphMat~\cite{sundaram15vldb-graphmat},
Galois~\cite{nguyen13sosp-galois}, Green-Marl~\cite{Hong12asplos},
Gemini~\cite{Zhu16gemni}, and Grazelle~\cite{Grossman2018}) with seven
algorithms consisting of PageRank (PR), Breadth-First Search (BFS),
Connected Components (CC) with synchronous label propagation, Single
Source Shortest Paths (SSSP) with frontier-based Bellman-Ford,
Collaborative Filtering (CF), Betweenness Centrality (BC), and PageRankDelta (PRDelta), running on
real-world graphs with different sizes and structures.  Our
experiments show that \graphit outperforms the next fastest of the
shared-memory frameworks on \numFastExp out of \numExp experiments by up to
4.8$\times$, and is never more than 43\% slower than the fastest
framework on the other experiments.  For each framework and DSL, we
show a heat map of slowdowns compared to the fastest of all seven
frameworks and DSLs in \Fig~\ref{fig:intro_comparisons}.  \graphit
does not introduce any new optimizations.  Instead, the DSL achieves
competitive or better performance compared to other frameworks by
generating efficient implementations of known combinations of
optimizations, and finding previously unexplored combinations by
searching through a much larger space of optimizations.  \graphit also
reduces the lines of code compared to the next fastest framework by up 
to an order of magnitude.

\punt{
Our current target users are programmers who are familiar with graph
optimizations. \graphit enables these programmers to quickly tune
the performance of their code without having to manually write each
combination of optimizations. Popular DSLs such as Halide~\cite{Ragan-Kelley:2013:HLC:2499370.2462176} have validated the practicality of this methodology when high performance matters. 
}


This paper makes the following contributions:

\begin{denseitemize}
\item A systematic analysis of the fundamental tradeoffs among
  locality, \work, and parallelism in graph optimizations
  (Section~\ref{sec:tradeoff}).
\item A high-level algorithm language that separates edge processing
  logic from edge traversal, synchronization, updated vertex tracking,
  and deduplication logic (Section~\ref{sec:algo}).
\item A new scheduling language that allows programmers to explore the
  tradeoff space by composing edge traversal, vertex data layout, 
  and program structure optimizations (Section~\ref{sec:schedule}).
\item A novel scheduling representation, the graph iteration space
  model, that can represent, combine and reason about the validity of
  various edge traversal optimizations
  (Section~\ref{sec:schedule_rep}).
\item A compiler that leverages program analyses on the algorithm
  language and an intermediate representation that encodes the graph
  iteration space to generate efficient and valid implementations for
  different combinations of optimizations. 
  (Section~\ref{sec:compiler}).
\item A comprehensive evaluation of \graphit that shows it is faster
  than the next fastest state-of-the-art framework on ~\numFastExp out of ~\numExp
  experiments by up to 4.8$\times$, and never more than 43\% slower
  than the fastest framework on the other experiments
  (Section~\ref{sec:eval}).
\end{denseitemize}

\section{Tradeoff Space and Optimizations}
\label{sec:tradeoff}

\algrenewcommand{\alglinenumber}[1]{\footnotesize#1}
\algdef{SE}[SUBALG]{Indent}{EndIndent}{}{\algorithmicend\ }%
\algtext*{Indent} \algtext*{EndIndent}
\algblockdefx[pfor]{ParFor}{EndParFor}[1] {\textbf{parallel
    for}~#1~\textbf{do}}{}  
\algtext*{EndParFor}

\begin{wrapfigure}[17]{R}{0.58\textwidth}
\begin{algorithmic}[1]
  \scriptsize
  \State Rank $= \{0, \ldots, 0\}$  \Comment{Length $V$ array}
  \State DeltaSum $= \{0.0, \ldots, 0.0\}$  \Comment{Length $V$ array}
  \State Delta $= \{1/V, \ldots, 1/V\}$  \Comment{Length $V$ array}
  \Procedure{PageRankDelta}{Graph $G$, $\alpha$,
    $\epsilon$} \State Frontier = $\{$ $G$.vertices
  $\}$ \label{line:init-frontier} \For {round
    $\in\{1,\ldots,$MaxIter$\}$} \State NextFrontier = $\{ \}$
  \ParFor{src : Frontier} \label {line:access_src} \For{dst :
    $G$.getOutNgh[src]} \label
  {line:access_dst} 
  \State AtomicAdd(DeltaSum[dst], Delta[src]$/$G.OutDegree[src]) \label{line:update_dst} \EndFor \EndParFor \ParFor
  {v : $G$.vertices} \label{line:start_delta_comp} \If {round $== 1 $}
  \State BaseScore $=(1.0 - \alpha)/V$ \State Delta[v]
  $= \alpha \cdot ($DeltaSum[v]$) +$ BaseScore \State Delta[v]
  $\mathrel{-}= 1/V$ \Else {} \State Delta[v]
  $= \alpha \cdot ($DeltaSum[v]$)$ \EndIf \State Rank[v]
  $\mathrel{+}=$ Delta[v] \label{line:end_delta_comp} \State
  DeltaSum[v] $=$ 0 \If
  {$\mid$Delta[v]$\mid > \epsilon \cdot
    $Rank[v]} \label{line:vertex_filtering} \State
  NextFrontier.add(v) \label{line:filling_frontier} \EndIf
  \EndParFor \label{line:end_vertex_filtering} \State Frontier $=$
  NextFrontier \EndFor \label{line:done} \EndProcedure
\end{algorithmic}
\caption{PageRankDelta (SparsePush).}
\label{alg:PageRankDelta}
\end{wrapfigure}

\punt{
In this section, we present the tradeoff space for optimizing graph
programs. 
}
While the effects of various optimizations are well-known
to expert programmers, we believe that we are the first to
characterize the optimization 
tradeoff space for graph optimizations. Our tradeoff space
includes three properties of graph programs---locality, \work, and
parallelism. \defn{Locality} refers to the amount of spatial and
temporal reuse in a program. Increasing the amount of reuse improves
locality due to increasing the number of cache hits. In a NUMA system,
locality also refers to the memory location relative to the
processor. Increasing the ratio of local memory accesses to remote
memory accesses improves locality.
\defn{\Work} is the inverse of the number of
instructions, where each instruction is weighted according to the
number of cycles that it takes to execute assuming a cold cache.\punt{\footnote{We note that if we fix a baseline sequential algorithm, this definition is compatible with the definition of \work in traditional parallel algorithm analysis.}} Reducing the number
of instructions improves \work.
  \defn{Parallelism} refers to the
relative amount of work that can be executed independently by
different processing units, which is often affected by the load
balance and synchronization among processing units. Increasing
parallelism improves performance by taking advantage of more
processing units, and helping to hide the latency of DRAM requests. We
use PageRankDelta (described in Section~\ref{subsec:tradeoff_prdelta} and shown in \Fig~\ref{alg:PageRankDelta})
as an example to illustrate the effects of various optimizations on
these three metrics. Each optimization can affect multiple properties
in the tradeoff space, either positively or negatively.  The complex
tradeoff space motivates the design of \graphit's scheduling language
and compiler, which can be used to easily search for points in the
tradeoff space that achieve high performance.

\subsection{PageRankDelta}
\label{subsec:tradeoff_prdelta}
\defn{PageRankDelta}~\cite{shun13ppopp-ligra} is a variant of the
standard PageRank algorithm~\cite{page99tr-pagerank} that computes the
importance of vertices in a graph. It maintains an array of ranks, and
on each iteration, updates the ranks of all vertices based on the
ranks of their neighbors weighted by their neighbors'
out-degrees. PageRankDelta speeds up the computation by updating only
the ranks of vertices whose ranks have changed significantly from the
previous iteration.

The pseudocode for PageRankDelta is shown in
Fig.~\ref{alg:PageRankDelta}, where $V$ is the number of vertices in
the graph, $0 \leq \alpha \leq 1$ is the damping factor that
determines how heavily to weight the neighbors' ranks during the
update, and $\epsilon \geq 0$ is a constant that determines whether a
vertex's rank has changed sufficiently.  For simplicity, we update the
ranks of vertices for MaxIter number of iterations, although the code
can easily be modified to terminate based on a convergence
criterion. The algorithm maintains the set of vertices whose ranks
(stored in the Rank array) have changed significantly from the
previous iteration in the variable Frontier (represented as a sparse array). We will refer to this as
the \emph{active set} of vertices, or the \emph{frontier}.  Initially
all vertices are in the active set (Line~\ref{line:init-frontier}). On
each iteration, each vertex in the active set sends its Delta (change
in Rank value) from the previous iteration to its out-neighbors by
incrementing the DeltaSum values of its neighbors
(Lines~\ref{line:access_src}--\ref{line:update_dst}). Since vertices
are processed in parallel, the updates to DeltaSum must be atomic.
Then in parallel, all vertices compute their own Delta and Rank values
based on their DeltaSum value and $\alpha$
(Lines~\ref{line:start_delta_comp}--\ref{line:end_delta_comp}).  
Delta is computed differently for the first iteration.  If the Delta
of the vertex is larger than $\epsilon$ times its Rank, then the
vertex is active for the next iteration and is added to the next
frontier
(Lines~\ref{line:vertex_filtering}--\ref{line:end_vertex_filtering}).

\subsection{Graph Optimizations}
\label{subsec:tradeoff_optimizations}
We describe the tradeoffs of the
optimizations listed in Table~\ref{table:opt-table} with
PageRankDelta (Table~\ref{table:related-table} describes which optimizations are supported by which frameworks and DSLs).
Table~\ref{table:opt-table} contains the effect of
optimizations relative to the baseline in
Fig.~\ref{alg:PageRankDelta}, which we refer to as \emph{SparsePush}.

\punt{ We refer to this optimization option as \emph{bitvector}, which
  is used in the algorithm shwon in
  Fig.~\ref{alg:PageRankDeltaPullBitvectorWithStructs}.  }

\begin{wraptable}[17]{R}{0.59\textwidth}
  \footnotesize
  \tabcolsep 3pt

\centering
\centering
\captionof{table}{\small Effect of optimizations on the different properties of the tradeoff space relative to the baseline SparsePush. For each property, $\uparrow$ means positive impact on performance, $\downarrow$ means negative impact on performance, $\updownarrow$ means it could increase or decrease depending on various factors (described in the text), and no entry means no effect on performance.
}
\begin{tabular}[!t]{c|c|c|c}
Optimization & Locality & \Work & Parallelism  \\
\hline\hline
DensePull & & $\updownarrow$ & $\updownarrow$ \\ 
DensePush & & $\updownarrow$ & $\uparrow$  \\
DensePull-SparsePush & & $\updownarrow$ & $\updownarrow$ \\ 
DensePush-SparsePush & & $\updownarrow$ & $\uparrow$\\
edge-aware-vertex-parallel & & $\downarrow$ & $\uparrow$\\
edge-parallel & & $\downarrow$ & $\uparrow$\\
bitvector & $\uparrow$ & $\downarrow$ &\\
vertex data layout & $\updownarrow$ & $\downarrow$ &\\

cache partitioning & $\uparrow$ & $\downarrow$ &\\
NUMA partitioning & $\uparrow$ & $\downarrow$ & $\downarrow$\\ 
kernel fusion & $\uparrow$ & $\downarrow$ &\\ \hline
\end{tabular}
\label{table:opt-table}
\end{wraptable}

\punt{
kernel split & & $\downarrow$ &\\
kernel fusion & $\uparrow$ & $\uparrow$ &\\
}


\myparagraph{Direction Optimization and Frontier Data Structure}
Lines~\ref{line:access_src}--\ref{line:update_dst} of SparsePush
iterate over the outgoing neighbors of each vertex, and update the
DeltaSum value of the neighbor.  \emph{DensePull}
(Fig.~\ref{alg:PageRankDeltaDense} (left)) is a different traversal
mode where each vertex iterates over its incoming neighbors that are
in the active set, and updates its own DeltaSum value.  DensePull
increases parallelism relative to SparsePush since it loops over all
vertices in the graph.  This increases work compared to SparsePush,
which only loops over vertices in the active set.  The update to the
DeltaSum array no longer requires atomics since an entry will not be
updated in parallel, and this reduces synchronization
overhead. Instead of doing random writes as in SparsePush, DensePull
does random reads and mostly sequential writes, which are cheaper. For
some algorithms (e.g., breadth-first search), the inner loop over the
in-neighbors in DensePull can exit early to reduce overall work.
Therefore, the overall number of edges traversed could increase or
decrease.  A detailed performance study of the two traversal methods
can be found in~\citet{Beamer-2012} and \citet{Besta2017}.  We can further use
\emph{bitvectors} instead of boolean arrays to keep track of vertices on the
frontier for the DensePull direction.  A dense frontier implemented
using a bitvector improves spatial locality, but requires extra
work to compress the boolean array.

\emph{DensePush} (Fig.~\ref{alg:PageRankDeltaDense} (right)) loops
through all vertices and checks if each one is on the frontier instead
of only looping over frontier vertices as in SparsePush.  While iterating over all
vertices reduces \work, this could be offset by not having to maintain
the frontier in a sparse format. Parallelism increases as there is more parallel work when looping over all vertices.

\begin{figure}[t]
  \hspace{-20pt}
  \begin{subfigure}[t]{.48\textwidth}
    \begin{algorithmic}[1]
      \scriptsize \ParFor{dst : $G$.vertices} \label
      {line:pull_access_dst} \For{src : $G$.getInNgh[dst]} \label
      {line:pull_access_src} \If{src $\in$ Frontier} \State
      DeltaSum[dst] $\mathrel{+}=$ Delta[src] $/$
      G.OutDegree[src] \label{line:pull_update_dst} \EndIf \EndFor
      \EndParFor
    \end{algorithmic}

  \end{subfigure}
  \hspace{-10pt}
  \begin{subfigure}[t]{.54\textwidth}
    \begin{algorithmic}[1]
      \scriptsize \ParFor{src : $G$.vertices} \label
      {line:dense_push_access_src} \If{src $\in$ Frontier} \label
      {line:dense_push_check_src} \For{dst :
        $G$.getOutNgh[src]} \label
      {line:dense_push_access_dst} 
                
      \State AtomicAdd(DeltaSum[dst],Delta[src] $/$
      G.OutDegree[src]) \label{line:pull_update_dst} \EndFor \EndIf
      \EndParFor
    \end{algorithmic}
  \end{subfigure}
  \hspace{-10pt}
  \caption{DensePull (left) and DensePush
    (right).}\label{alg:PageRankDeltaDense}
\end{figure}

Hybrid traversal modes (\emph{DensePull-SparsePush} and
\emph{DensePush-SparsePush}) use different directions (SparsePush,
DensePull, and DensePush) in different iterations based on the size of
the active set to improve \work~\cite{Beamer-2012,shun13ppopp-ligra}.
In PageRankDelta, the number of vertices in the frontier gradually
decreases as the ranks of vertices converge. In the early iterations,
DensePull is preferred due to lower synchronization overheads and
avoiding random writes. As the frontier shrinks, SparsePush is
preferred due to the fewer number of vertices that need to be
traversed. \emph{DensePull-SparsePush} computes the sum of out-degrees
of the frontier vertices and uses DensePull if the sum is above some
threshold, and SparsePush otherwise. However, computing the sum of
out-degrees of the vertices in the active set in every iteration
incurs significant overhead if one direction is always better than the
other.

\punt{As shown in Table~\ref{table:related-table}, }

\graphit is able to
support all of these traversal directions whereas existing frameworks
only support a subset of them. \graphit also supports both 
bitvectors and boolean arrays for the frontier representation in dense
traversals, as well as the sparse array representation for sparse traversals.

\punt{For some algorithms, DensePush can perform better than
  DensePull, which leads to another hybrid version
  \emph{DensePush-SparsePush}.

  , and disabling early breaking of the inner loop for DensePul

}

\myparagraph{Parallelization}
For each traversal mode, there are different methods for
parallelization. The parallelization shown in
Fig.~\ref{alg:PageRankDelta} and \ref{alg:PageRankDeltaDense}
processes the vertices in parallel.
This approach (\emph{vertex-parallel}) works well on algorithms and inputs where
the workload of each vertex is similar. However, if the degree
distribution is skewed and the workload of each vertex is proportional to
the number of incident edges, this approach can lead to significant
load imbalance.  For these workloads, an edge-aware vertex-parallel
scheme (\emph{edge-aware-vertex-parallel}) can be more effective.
This approach breaks up the vertices into a number of vertex chunks,
where each chunk has approximately the same number of edges. However,
this scheme reduces \work due to having to compute the sum of degrees
of vertices in each chunk.  Finally, we can parallelize across all
edges, instead of just vertices, by parallelizing the inner loop of
the edge-traversal code computing DeltaSum in
Fig.~\ref{alg:PageRankDelta} and~\ref{alg:PageRankDeltaDense}. This
method (\emph{edge-parallel}) improves parallelism but reduces \work
due to the overhead of work-stealing in the inner loop and atomic
updates needed for synchronization.  For graphs with a regular degree
distribution, using static parallelism instead of work-stealing
parallelism can sometimes reduce runtime overhead and improve performance.
\graphit supports all three modes of parallelism while existing frameworks only support one or two. 

\punt{ More sophisticated schemes for Edge-parallel can be
  implemented, such as only parallelizing across edges for high-degree
  vertices, and for DensePull parallelizing across edges in chunks and
  checking the early break condition after each chunk.}

\punt{
  \begin{wrapfigure}[5]{L}{0.65\textwidth}
    \begin{algorithmic}[1]
      \scriptsize \State Bitvec =
      ConvertToBitvec(Frontier)\label{line:pull_opt_setup_bitvec}
      \ParFor{dst : $G$.vertices} \label {line:pull_access_dst}
      \For{src : $G$.getInNgh[dst]} \label {line:pull_access_src}
      \If{Bitvec.getBit(src)} \label{line:pull_opt_use_bitvec} \State
      DeltaSum[dst] $\mathrel{+}=$
      FusedStruct[src].Delta$/$FusedStruct[src].OutDegree\label{line:pull_opt_update_dst}
      \EndIf \EndFor \EndParFor
    \end{algorithmic}
    \caption{DensePull with Vertex Data Layout Optimizations.}
    \label{alg:PageRankDeltaPullBitvectorWithStructs}
  \end{wrapfigure}
}

\myparagraph{Cache Partitioning} Cache partitioning tries to keep
random accesses within the last level cache (LLC) to improve
locality. This optimization first partitions the vertices into $p$
segments ($V_0, V_1, \ldots, V_{p-1}$), which correspond to the range
of source \textsf{vertexset}s in the pull mode or the destination \textsf{vertexset}s in
the push mode for each Segmented Subgraph (SSG).  For the pull mode,
incoming edges ($src$, $dst$) are assigned to $SSG_i$ if $src \in V_i$, and sorted by $dst$.
For the push mode, outgoing edges ($src$, $dst$) are assigned to
$SSG_i$ if $dst\in V_i$, and sorted by $src$.
Each SSG is processed before moving on to the next.
$V_i$ controls the range of random memory accesses through segmenting the original graph.  If we
fit $V_i$ into LLC, we can significantly reduce the number of random
DRAM accesses when processing each SSG. Cache partitioning improves
locality but sacrifices \work due to vertex data replication from graph
partitioning and merging partial
results~\cite{Yunming2017,Beamer2017,Nishtala2007}. Fine-tuning the number of SSGs
can reduce this overhead. Most existing frameworks do not support cache partitioning.

\punt{ Non-uniform memory access is a multiprocessing design where the
  memory access time from a processor depends on the memory location
  relative to the processor.

  Since inter-socket latencies can be 2 to 7.5 times longer than
  intra-socket latencies depending on the architecture ~\cite{Sync},
  NUMA-aware graph processing that minimizes remote memory access
  improves performance.

  requires graph partitioning which can be done in the same way as the
  cache optimization.

  Whether to use NUMA optimization or to simply allocate memory in an
  interleaved fashion depends on the particular application and the
  graph.

}

\myparagraph{NUMA Optimizations} NUMA partitioning improves locality by
minimizing slow inter-socket memory
accesses~\cite{zhang15ppopp-numa-polymer, Sun2017, Zhu16gemni}.  
 This optimization partitions the graph into a set of Segmented Subgraphs (SSGs) in the
same way as cache partitioning in order to limit the range of random memory
access.  While the cache partitioning optimization processes one SSG at a time across
all sockets, NUMA partitioning executes multiple SSGs in parallel on
different sockets.  Each SSG and the threads responsible for
processing the subgraph are bound to the same NUMA socket. The
intermediate results collected on each socket are merged at the
end of each iteration. As with cache partitioning, NUMA partitioning improves locality
but reduces \work due to vertex data replication from graph
partitioning and the additional merge phase.  Parallelism might also
decrease for highly skewed graphs due to workload imbalance among
SSGs~\cite{Sun2017}. For algorithms with performance bottlenecked on 
load imbalance instead of inter-socket memory accesses,
simply using an interleaved allocation across sockets
can result in better performance. \graphit and a subset of existing frameworks support NUMA optimizations.

\punt{ As shown in Table~\ref{table:related-table},}

\myparagraph{Vertex Data Layout Optimizations} The layout of vertex
data can significantly affect the locality of memory accesses.  Random
accesses to the same index of two separate arrays (e.g., the Delta and
OutDegree arrays in PageRankDelta) can be changed into a single random
access to an array of structs to improve spatial locality.  However,
grouping together fields that are not always accessed together expands
the working set and hurts the locality of the data structures. Vertex
data layout optimizations reduce work-efficiency due to the extra
overhead for reorganizing the data.  \graphit supports both arrays of
structs and structs of arrays. \punt{ as well as the automatic transformation
between the two. Existing frameworks use only one of the two
representations (Table~\ref{table:related-table}).}

\updated{
\myparagraph{Program Structure Optimizations}
When two graph kernels have the same traversal pattern (they process
the same vertices/edges on each iteration), we can fuse
 together the edge traversals and
transform their data structures into an array of structs.
 We refer to this optimization
as \emph{kernel fusion}. This improves spatial locality by enabling
the program to access the fields of the two kernels together when
traversing the edges. Additional work is incurred for performing the
AoS-SoA optimization, but this is usually small compared to the
rest of the algorithm. 
} 

\section{Algorithm Language}
\label{sec:algo}

\punt{
For example,
we can easily come up with more than 20 optimized
implementations for PageRankDelta (at least 5 directions, 3 parallelization strategies,
vertex data layout optimization, cache optimization, and NUMA optimization),
but it would be tedious to hand-write
all the different implementations.  

Section~\ref{sec:tradeoff} described the
challenge of finding the right combination of optimizations for
graph algorithms, as performance depends on many factors and there 
are a large number of combinations of optimizations.
\graphit leverages an
algorithm representation and a separate scheduling language to compose together
graph optimizations, enabling programmers to easily navigate the
complex performance tradeoff space.
}

\begin{figure}
\begin{lstlisting} [language=graphit,escapechar=|]
element Vertex end |\label{line:vertex_element}|
element Edge end |\label{line:edge_element}|
const edges : edgeset{Edge}(Vertex,Vertex) = load(argv[1]); |\label{line:edgeset_def}|
const vertices : vertexset{Vertex} = edges.getVertices(); |\label{line:vertexset_def}|
const damp : double = 0.85;
const base_score : double = (1.0 - damp)/vertices.size();
const epsilon : double = 0.1;
const OutDegree : vector{Vertex}(int) = edges.getOutDegrees(); |\label{line:outdegree_def}|
Rank : vector{Vertex}(double) = 0; |\label{line:rank_def}|
DeltaSum : vector{Vertex}(double) = 0.0;
Delta : vector{Vertex}(double) = 1.0/vertices.size(); |\label{line:delta_def}|
func updateEdge(src : Vertex, dst : Vertex) |\label{line:graphit_edge_processing_start}|
    DeltaSum[dst] += Delta[src]/OutDegree[src]; |\label{line:graphit_edge_processing}|
end |\label{line:graphit_edge_processing_end}|
func updateVertexFirstRound(v : Vertex) -> output : bool
    Delta[v] = damp * (DeltaSum[v]) + base_score;
    Rank[v] += Delta[v];
    Delta[v] = Delta[v] - 1.0/vertices.size();
    output = fabs(Delta[v] > epsilon*Rank[v]);
    DeltaSum[v] = 0;
end
func updateVertex(v : Vertex) -> output : bool
   Delta[v] = DeltaSum[v] * damp;
   Rank[v] += Delta[v];
   DeltaSum[v] = 0;
   output = fabs(Delta[v]) > epsilon * Rank[v];
end
func main()
    var V : int = vertices.size();
    var Frontier : vertexset{Vertex} = new vertexset{Vertex}(V);
    for i in 1:maxIters |\label{line:for_loop}|
        #s1# edges.from(frontier).apply(updateEdge); |\label{line:edgeset_operators}|
        if i == 1
            Frontier = vertices.filter(updateVertexFirstRound); |\label{line:vertexset_filter_round_one}|
        else
            Frontier = vertices.filter(updateVertex); |\label{line:vertexset_filter}|
        end
    end
end
\end{lstlisting}
\caption{\graphit code for PageRankDelta.}
\label{fig:code:pagerankdelta}
\end{figure}

\graphit leverages an
algorithm and a scheduling language to compose
graph optimizations, enabling programmers to easily navigate the
complex performance tradeoff space described in Section~\ref{sec:tradeoff}.
The algorithm language can express a
variety of algorithms, while exposing opportunities for optimizations. 
 We use PageRankDelta
(\graphit code shown in \Fig~\ref{fig:code:pagerankdelta}) to showcase the language.

\punt{
The abstract data model enables vertex data layout
optimizations. The set operators separate the edge traversal logic
from the update logic, allowing for direction and
parallelization optimizations during edge traversal.
}

\punt{

\begin{table}[t]
\footnotesize
 \begin{tabular}{p{2cm}p{4cm}}
\textbf{Data Type}  & \textbf{Example} \\
 \textsf{element} & \lstinline|element Vertex end;|\\
 \textsf{vector} & \lstinline|rank : vector (Vertex)[float] = 0;|\\
 \textsf{vertexset} & \lstinline|var vertices : vertexset {Vertex}; | \\
\textsf{edgeset} & \lstinline|var edges : edgeset {Edge}(Vertex, Vertex); |\\
\noalign{\smallskip}
  \end{tabular}
  \caption{\graphit Data Types}
  \label{table:data_types}
\end{table}
}

\subsection{Data Model}
\graphit's data model consists of \textsf{element}s, \textsf{vertexset}s and
\textsf{edgeset}s, and vertex and edge data.  The programmer first defines vertex
and edge \textsf{element} types (\textsf{Vertex} and \textsf{Edge} on Lines~\ref{line:vertex_element}--\ref{line:edge_element} of
\Fig~\ref{fig:code:pagerankdelta}). \graphit supports multiple types
of user-defined vertices and edges, which is important for
algorithms that work on multiple graphs.  After defining \textsf{element} types,
the programmer can construct \textsf{vertexset}s and \textsf{edgeset}s.
Lines~\ref{line:edgeset_def}--\ref{line:vertexset_def} of \Fig~\ref{fig:code:pagerankdelta} show the
definitions of an \textsf{edgeset}, \textit{edges}, and \textsf{vertexset},
\textit{vertices}.  Each \textsf{element} of the \textsf{edgeset} is of \textsf{Edge} type
(specified between ``\{~\}''), and the source and destination of the edge
is of \textsf{Vertex} type (specified between ``$(\ )$'').  The \textsf{edgeset} declaration
supports edges with different types of source and destination vertices (e.g., in a bipartite
graph). \textit{vertices} uses the {\stt{getVertices}} method on the \textsf{edgeset} to
obtain the union of source and
destination vertices of \textit{edges}.  Data for vertices and edges are
defined as vectors associated with an \textsf{element} type denoted using the \{ \}
syntax (Lines~\ref{line:outdegree_def}--\ref{line:delta_def}).
\punt{ 
The
abstract sets implemented as array of vertices, boolean arrays,
or bitvectors. Vectors can be array of structs, struct of arrays,
or dictionaries.
}

\subsection{Language Constructs and Operators}
The language constructs of \graphit 
(shown in Table~\ref{table:set_api}) separate edge processing logic
from edge traversal, edge filtering
(\from, \too, \srcFilter, and \dstFilter), atomic synchronization, and modified vertex deduplication and tracking logic
(\apply{} and \applyModified).
This separation enables the
compiler to represent the algorithm at a high level, exposing
 opportunities for edge traversal and vertex data layout
optimizations. Moreover, it frees the programmer from specifying low-level implementation details, such as synchronization and deduplication logic.

\begin{table}[t]
 \footnotesize
   \caption{\textsf{Vertexset} and \textsf{Edgeset} API. \ftt{disable\_deduplication} is an optional parameter.}
  \label{table:set_api}
 \begin{tabular}{p{4.4cm}p{1.8cm}p{6.8cm}}
 \textbf{Set Operators}  & \textbf{Return Type} & \textbf{Description} \\ \hline
 \ftt{size()} & \textsf{int} & Returns the size of the set. \\
 \textbf{\textsf{Vertexset} operators} & \\ \hline
 { \ftt{filter(func f)}} & \textsf{vertexset} & Filters out vertices where \texttt{f}(vertex) returns true.\\
 { \ftt{apply(func f)}} & none & Applies \texttt{f}(vertex) to every vertex. \\
 \textbf{\textsf{Edgeset} operators} & \\	 \hline
 { \ftt{from(\textsf{vertexset} vset)}} & \textsf{edgeset} & Filters out edges whose source vertex is in the input \textsf{vertexset}.\\
 { \ftt{to(\textsf{vertexset} vset)}} & \textsf{edgeset} & Filters out edges whose destination vertex is in the input \textsf{vertexset}.\\
  { \ftt{filter(func f)}} & \textsf{edgeset} & Filters out edges where \texttt{f}(edge) returns true.\\
 { \ftt{srcFilter(func f)}} & \textsf{edgeset} & Filters out edges where \texttt{f}(source) returns true.\\
 { \ftt{dstFilter(func f)}} & \textsf{edgeset} & Filters out edges where \texttt{f}(destination) returns true.\\
 { \ftt{apply(func f)}} & none & Applies \texttt{f}(source, destination) to every edge. \\
 { \ftt{applyModified(func f, \newline vector vec, \newline [bool disable\textunderscore deduplication])}} & \textsf{vertexset} & Applies \texttt{f}(source, destination) to every edge. Returns a \textsf{vertexset} that contains destination vertices whose entry in the vector vec has been modified in \texttt{f}.  The programmer can optionally disable deduplication within modified vertices. Deduplication is enabled by default.\\ \hline
\noalign{\smallskip}
  \end{tabular}
\end{table}

In the \graphit code for PageRankDelta (\Fig~\ref{fig:code:pagerankdelta}), the \from{}
operator (Line~\ref{line:edgeset_operators}) ensures
that only edges whose source vertex is in the frontier are traversed,
and the \apply{} operator uses the {\stt{updateEdge}} function on the selected
edges to compute DeltaSum, corresponding to
Lines~\ref{line:access_src}--\ref{line:update_dst} of
Algorithm~\ref{alg:PageRankDelta}. 
This separation enables the
compiler to generate complex code for different traversal modes and
parallelization optimizations, while inserting appropriate data access
and synchronization instructions for the \stt{updateEdge} function.
{\small \textbf{\texttt{\#s1\#}}} is a label used in the 
scheduling language (explained in Section~\ref{sec:schedule}).
Lines~\ref{line:vertexset_filter_round_one} and
\ref{line:vertexset_filter} of \Fig~\ref{fig:code:pagerankdelta}
compute the updated Delta and Rank values by applying
updateVertexFirstRound and updateVertex functions on every vertex. 
Vertices with Delta greater than epsilon of their Rank are
returned as the next frontier, corresponding to
Lines~\ref{line:start_delta_comp}--\ref{line:end_vertex_filtering} of
Algorithm~\ref{alg:PageRankDelta}.  As shown in
Table~\ref{table:set_api}, \graphit provides various operators
on \textsf{vertexset}s and \textsf{edgeset}s to express graph algorithms with
different traversal and update logic. 
The \applyModified{} 
operator tracks which
vertices have been updated during the edge traversal and outputs a
\textsf{vertexset} containing just those vertices. By default, \applyModified{} 
ensures that each vertex is added only once to the output
\textsf{vertexset}. However, the programmer can optionally disable
deduplication for algorithms that are guaranteed to insert each vertex
only once (e.g., BFS) for better performance. 

\begin{wrapfigure}[14]{R}{0.65\textwidth}
\begin{lstlisting} [language=c++,escapechar=|]
template <class vertex>
struct PR_F {
  vertex* V;
  double* Delta, *nghSum;
  PR_F(vertex* _V, double* _Delta, double* _nghSum) : 
    V(_V), Delta(_Delta), nghSum(_nghSum) {}
  inline bool update(uintE s, uintE d){
    double oldVal = nghSum[d]; |\label{line:ligra_edge_processing_1}|
    nghSum[d] += Delta[s]/V[s].getOutDegree(); |\label{line:ligra_edge_processing_2}|
    return oldVal == 0;} |\label{line:ligra_modification_1}|
  inline bool updateAtomic (uintE s, uintE d) {
    volatile double oldV, newV; |\label{line:ligra_sync_begin}|
    do { oldV = nghSum[d]; newV = oldV + Delta[s]/V[s].getOutDegree();
    } while(!CAS(&nghSum[d],oldV,newV));|\label{line:ligra_sync_end}|
    return oldV == 0.0;} |\label{line:ligra_modification_2}|
  inline bool cond (uintE d) { return cond_true(d); }}; |\label{line:ligra_edge_filtering}|
\end{lstlisting}
\caption{Ligra's PageRankDelta edge update function, corresponding to Lines~\ref{line:graphit_edge_processing_start}--\ref{line:graphit_edge_processing_end} of \Fig~\ref{fig:code:pagerankdelta} in \graphit's PageRankDelta example.  }
\label{fig:PRD_ligra}
\end{wrapfigure}

We demonstrate how \graphit simplifies
the expression of the algorithm 
by showing Ligra's implementation of the edge update function in \Fig~\ref{fig:PRD_ligra} (note that the 16 lines of Ligra code shown correspond to only 3 lines in \graphit's implementation in Fig.~\ref{fig:code:pagerankdelta}).
 Ligra requires the programmer to specify edge processing (Lines~\ref{line:ligra_edge_processing_1}--\ref{line:ligra_edge_processing_2}, \ref{line:ligra_sync_begin}--\ref{line:ligra_sync_end}),  
 edge filtering (Line~\ref{line:ligra_edge_filtering}), 
 deduplication and modification tracking (Lines~\ref{line:ligra_modification_1} and~\ref{line:ligra_modification_2}),
and synchronization logic (Lines~\ref{line:ligra_sync_begin}--\ref{line:ligra_sync_end}). 
\graphit only requires the programmer to specify the edge processing logic in this case. 

\punt{The full Ligra implementation is shown in the appendix.}

\punt{
Using PageRankDelta as an example, one might also want to keep track of which vertices have received DeltaSum updates at each iteration.
We can easily get the updated \textsf{vertexset} by replacing Line~\ref{line:edgeset_operators} of \Fig~\ref{fig:code:pagerankdelta} with the following code:

\begin{lstlisting}[language = graphit]
var updated_vertices : vertexset{Vertex} =
	edges.from(Frontier).applyModified(updateEdge, DeltaSum);
\end{lstlisting}
}

\graphit also supports traditional control flow constructs such as {\stt{for}}, {\stt{while}}, and {\stt{if}} for expressing fixed iteration loops, loops until convergence, and conditional control flow. After setting up a new \textsf{vertexset} called Frontier, Line~\ref{line:for_loop} in \Fig~\ref{fig:code:pagerankdelta} uses a {\stt{for}} loop to iterate maxIters times. An alternative implementation could use a while loop that iterates until the ranks of all vertices stabilize. 


\punt{ The \from and \too operators are evaluated lazily when combined with an \apply operator. \textit{for} and \textit{while} loop can only be used in iterative applications. They cannot be used to traverse \textsf{vertexset}s or \textsf{edgeset}s. This is enforced by making sure that graph traversal can be done only through the \apply operator. \textit{if} statements are used to express \emph{control flow} in the iterative graph applications. They cannot be used inside \apply functions as this would prevent the compiler from performing many performance optimizations. Instead, filtering can be expressed through a combination of \from, \too, and \textit{filter} operators.
}

\section{Scheduling Language}
\label{sec:schedule}

After specifying the algorithm using the language described in
Section~\ref{sec:algo}, programmers can explore different combinations
of optimizations using \graphit's scheduling language.  In this section, we
describe the design of the scheduling language functions and
demonstrate how they work with PageRankDelta.

\subsection{Scheduling Language}
\label{subsec:schedule_lang}


We use labels ({\small \textbf{\texttt{\#label\#}}}) in algorithm
specifications to identify the
statements on which optimizations apply.
 Programmers can assign a label on
the left side of a statement and later reference it in the scheduling
language. 
\Fig~\ref{lst:pagerankdelta_schedule} shows 
a simple schedule for the PageRankDelta implementation
in \Fig~\ref{fig:code:pagerankdelta}. The programmer adds
label \textbf{s1} to the \textsf{edgeset} operation statement. After the
\scheduleKeyword{} keyword, the programmer can make a series of calls to 
scheduling functions.

\begin{wrapfigure}{r}{0.6\textwidth}

\begin{lstlisting} [firstnumber=30, language=graphit,escapechar=|]
		...
		for i in 1:maxIters
			#s1# edges.from(frontier).apply(updateEdge); |\Suppressnumber|
			...|\Reactivatenumber{38}|
		end |\Suppressnumber|
...|\Reactivatenumber{41}|
schedule:
program->configApplyDirection("s1", "DensePull-SparsePush");
\end{lstlisting}
\caption{Scheduling PageRankDelta.}\label{lst:pagerankdelta_schedule}
\end{wrapfigure}

\begin{table}[t]
\footnotesize
  \caption{\graphit Scheduling Language functions. The default option for an operator is shown in bold. Optional arguments are shown in [ ]. If the optional direction argument is not specified, the configuration is applied to all relevant directions. We use a default grain size of 256 for parallelization.}
 \begin{tabular}{p{6.3cm}p{6.9cm}}
\textbf{Apply Scheduling Functions}  & \textbf{Descriptions} \\ \hline
 \ftt{program->configApplyDirection(label, config);} & Config options: \textbf{SparsePush}, DensePush, DensePull, DensePull-SparsePush, DensePush-SparsePush\\ \hline
 \ftt{program->configApplyParallelization(label, config, [grainSize], [direction]);} & Config options: \textbf{serial}, dynamic-vertex-parallel, static-vertex-parallel, edge-aware-dynamic-vertex-parallel, edge-parallel \\ \hline
 \ftt{program->configApplyDenseVertexSet(label, \newline config, [vertexset], [direction])} & \textsf{Vertexset} options: \textbf{both}, src-vertexset, dst-vertexset \newline Config Options: \textbf{bool-array}, bitvector \\ \hline
 \ftt{program->configApplyNumSSG(label, config, \newline numSegments, [direction]);} & Config options: \textbf{fixed-vertex-count}  or edge-aware-vertex-count\\ \hline
 \ftt{program->configApplyNUMA(label, config, \newline [direction]);} &  Config options: \textbf{serial}, static-parallel, dynamic-parallel \\ \hline
 \ftt{program->fuseFields(\{vect1, vect2, ...\})} & Fuses multiple arrays into a single array of structs.\\ \hline
 \ftt{program->fuseForLoop(label1, la bel2, \newline fused\_label)} & Fuses together multiple loops. \\ \hline
 \ftt{program->fuseApplyFunctions(label1, label2, \newline fused\_func)} & Fuses together two edgeset apply operators. The fused apply operator replaces the first operator.\\ \hline
\noalign{\smallskip}
  \end{tabular}
  \label{table:schedule_api}
\end{table}

We designed \graphit's scheduling language functions (shown in Table~\ref{table:schedule_api})
to allow programmers to compose together edge traversal direction, frontier data structure,
parallelization, cache, NUMA, vertex data layout, and program structure optimizations discussed in Section~\ref{sec:tradeoff}. 
The \stt{configApplyDirection} function allows programmers to configure directions
used for traversal. The programmer can use the \stt{configDenseVertexSet} function 
to switch between
bitvector and boolean array for source and destination \textsf{vertexset}s.
The \stt{configApplyNumSSG}
function configures the number of segmented subgraphs and how the subgraphs are partitioned (fixed-vertex-count and edge-aware-vertex-count).
Setting the right number of segments and
 partitioning configuration allows random accesses to be restricted to
a NUMA node or last level cache with
balanced load as described in Section~\ref{sec:tradeoff}.
\stt{configApplyNUMA} configures the segmented subgraphs
to be executed in parallel with static or dynamic NUMA node assignment
(static-parallel and dynamic-parallel), ensuring the random memory
accesses are restricted to the local NUMA node, while maintaining good parallel scalability.
 Finally, vertex data vectors can be fused together into an 
 array of structs with \stt{fuseFields}.

To compose together different optimizations,
 the programmer first chooses a direction for traversal.
Then the programmer can use the other scheduling functions to pick one option for the parallelization,
graph partitioning, NUMA, and dense \textsf{vertexset} optimizations for the current direction.
The programmer can configure each direction separately using the optional direction
argument for hybrid directions (DensePush-SparsePush or DensePull-SparsePush).
If no direction argument is specified,
then the configuration applies to both directions. 

\subsection {Scheduling PageRankDelta}
\label{subsec:order_data_layout}

\begin{figure*}[t]
\centering
\includegraphics[width=\textwidth]{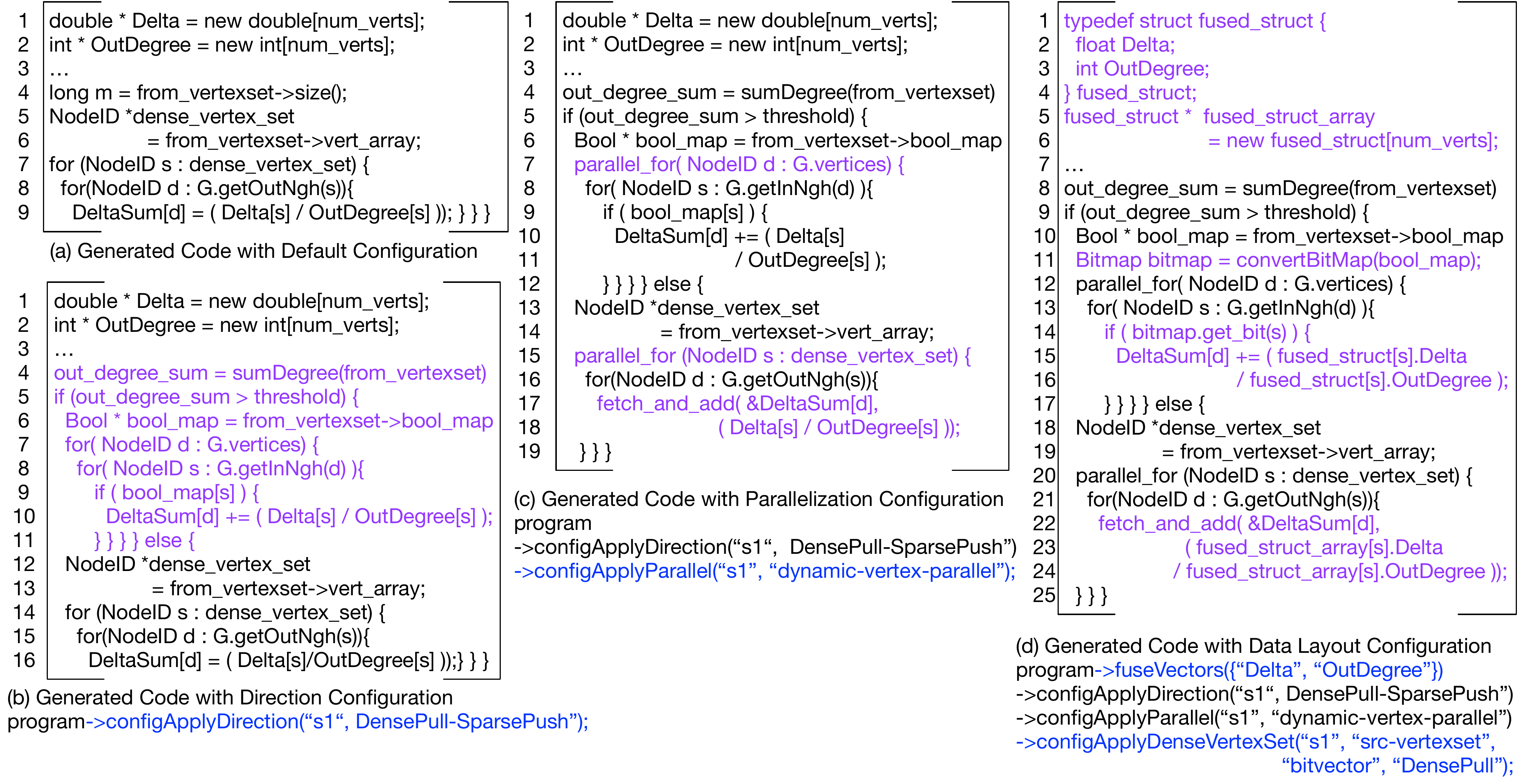}
\caption{Each subfigure shows pseudocode generated from applying the schedule in the caption to the \graphit PageRankDelta code with labels  from \Fig~\ref{fig:code:pagerankdelta} and \Fig~\ref{lst:pagerankdelta_schedule}. The options in the caption highlighted in blue are newly added scheduling commands relative to the previous subfigure and the code highlighted in purple is pseudocode updated due to the new schedules.
  }
\label{fig:prdelta_generated}
\end{figure*}

\Fig~\ref{fig:prdelta_generated} shows
different schedules for PageRankDelta.
\Fig~\ref{fig:prdelta_generated}(a) starts with the pseudocode
generated from the default schedule that performs a serial \SparsePush
traversal.  \Fig~\ref{fig:prdelta_generated}(b) adds hybrid
traversal code that first computes the sum of out-degrees and uses it to
determine whether to do a DensePull or a SparsePush traversal. This
allows the implementation to pick the traversal mode that
minimizes the number of edges that need to be traversed, improving \work.  \Fig~\ref{fig:prdelta_generated}(c) adds
dynamic-vertex-parallelism to both directions in the generated code 
by parallelizing the loops and inserting
synchronization code for SparsePush.  Finally,
 \Fig~\ref{fig:prdelta_generated}(d) adds vertex data layout and bitvector
optimizations. Fusing together the vectors Delta and OutDegree 
with the \stt{fuseFields} function improves
spatial locality of memory accesses since the two vectors are always accessed
together.  This optimization changes the declaration and
access points for the arrays. Finally, for the DensePull direction, 
the source \textsf{vertexset} specified in \from{}
can be dynamically compressed into a bitvector to reduce the
working set size, further improving spatial locality.

\subsection{Scheduling Program Structure Optimizations}
\label{subsec:structural}

To support program structure optimizations,
we introduce scoped  labels, which
allow labels to function even after complex program
transformations, and scheduling functions for fusing together loops and edgeset \apply{} operators. 
\Fig~\ref{lst:loop_fusion} shows two iterative edgeset \apply{} operators (Lines ~\ref{line:fusion_first_apply} and ~\ref{line:fusion_second_apply}) that can 
be fused together into a single iterative edgeset \apply{} operator. 
\graphit first performs
loop fusion, creating a new loop (\textbf{l3}), and destroying the two
old loops (\textbf{l1} and \textbf{l2}). Now, it would be difficult if
we wanted to schedule the first edgeset \apply{} operator in the
\textbf{l3} loop as the original loops \textbf{l1} and \textbf{l2} have
been removed from the program. Since both edgeset \apply{} operators have
\textbf{s1} as their label, it is hard to identify them individually. 
To address this, we introduce scoping to labels. The two  \apply{}
operators will obtain labels \textbf{l1:s1} and \textbf{l2:s1}, respectively.  

\begin{wrapfigure}{r}{0.54\textwidth}
   \begin{lstlisting} [language=graphit, escapechar=|]
#l1# for i in 1:10 
     #s1# edges.apply(func1); |\label{line:fusion_first_apply}|
end
#l2# for i in 1:10
     #s1# edges.apply(func2); |\label{line:fusion_second_apply}|
end
schedule:
program->fuseForLoop("l1", "l2", "l3")
->fuseApplyFunctions("l3:l1:s1", "l3:l2:s1", "fusedFunc")
->configApplyDirection("l3:l1:s1", "DensePull");
   \end{lstlisting}
   \vspace{-2pt}
   \caption{GraphIt loop and function fusion}\label{lst:loop_fusion}
   \vspace{-2pt}
\end{wrapfigure}

We also need a
name node, which enforces a named scope for the label. 
Loops \textbf{l1} and \textbf{l2} are replaced with name
nodes with labels \textbf{l1} and \textbf{l2}, respectively. The resulting pseudocode is shown in \Fig~\ref{lst:intermediate}.
This enables the user to reference
the first edgeset \apply{} as \textbf{l3:l1:s1} and the second edgeset \apply{} as \textbf{l3:l2:s1}.
After the loops are fused together, 
we can use \fuseApplyFunctions{} to create a
new edgeset \apply{} to replace the \textbf{l3:l1:s1} 
statement, which can be further configured (\Fig~\ref{lst:after_fusion}). The new edgeset \apply{} function, \stt{fusedFunc}, concatenates 
the statements in the original functions,
 \stt{func1} and \stt{func2}. 
In Section~\ref{sec:eval}, we show that fusion of multiple
iterative kernels with similar traversal patterns (Eigenvector Centrality and PageRank), and the vertex data vectors they access boosts the
performance of the application by up to 60$\%$.

\begin{figure}[h]
\begin{minipage}{.48\textwidth}
\begin{lstlisting} [language=graphit,escapechar=|]
#l3# for i in 1:10
    #l1# namenode |\label{line:name_node}|
         #s1# edges.apply(func1); 
    end
    #l2# namenode
         #s1# edges.apply(func2);
    end
end
\end{lstlisting}
\caption{Pseudocode after loop fusion} \label{lst:intermediate}
\end{minipage}%
\begin{minipage}{.48\textwidth}
\begin{lstlisting} [language=graphit,escapechar=|]
#l3# for i in 1:10
    #l1# namenode |\label{line:name_node}|
         #s1# edges.apply(fused_func); 
    end
end
\end{lstlisting}
\caption{Pseudocode after function fusion}\label{lst:after_fusion}
\end{minipage}
\end{figure}

\section{Scheduling Representation}
\label{sec:schedule_rep}

\updated{
\begin{figure}[t]
\centering
\includegraphics[width=0.95\textwidth]{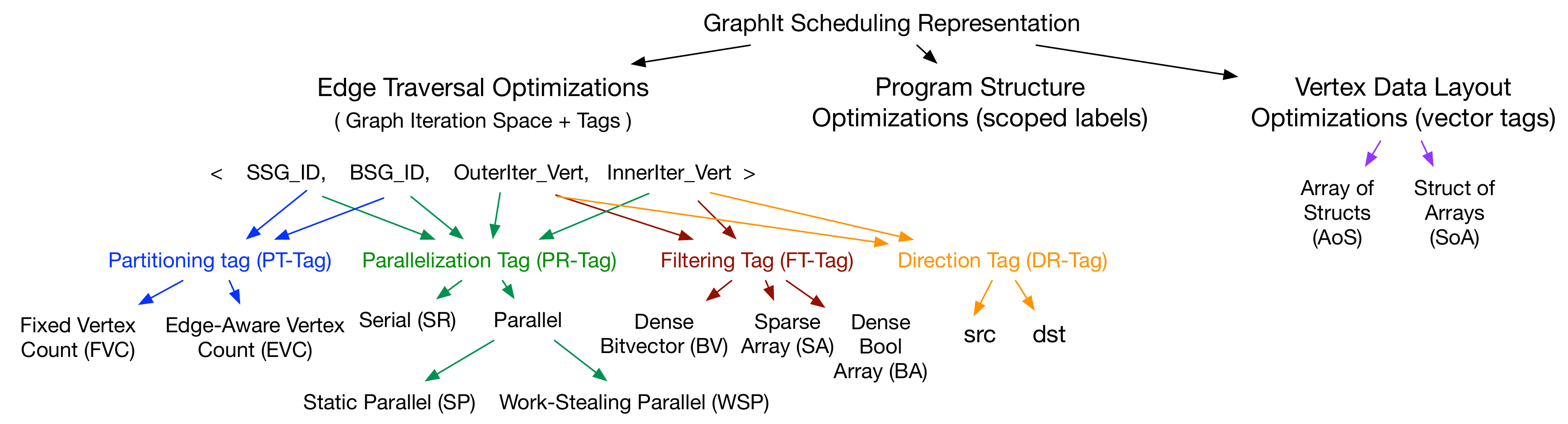}
\caption{GraphIt's scheduling representation for edge traversal, vertex data layout, and program structure optimizations. The tags of the graph iteration space represent the direction 
and performance optimization choices for each vertex 
data vector and each dimension of the graph iteration space.}
\label{fig:GIS_supported}
\end{figure}

The schedules for an optimized PageRankDelta implementation
 become even more complex than those shown 
in \Fig~\ref{fig:prdelta_generated}(d) as we further 
combine NUMA and cache optimizations. 
It is challenging to reason about 
the validity of, and to generate code for,
combinations of optimizations.
\graphit relies on multiple scheduling representations, specifically
the graph iteration space, the vertex data vector tags, and the scoped labels, to model combinations 
of edge traversal, vertex data layout, and program structure optimizations. \Fig~\ref{fig:GIS_supported} shows the full space of 
optimizations.

\begin{figure}[t]
\includegraphics[width=\textwidth]{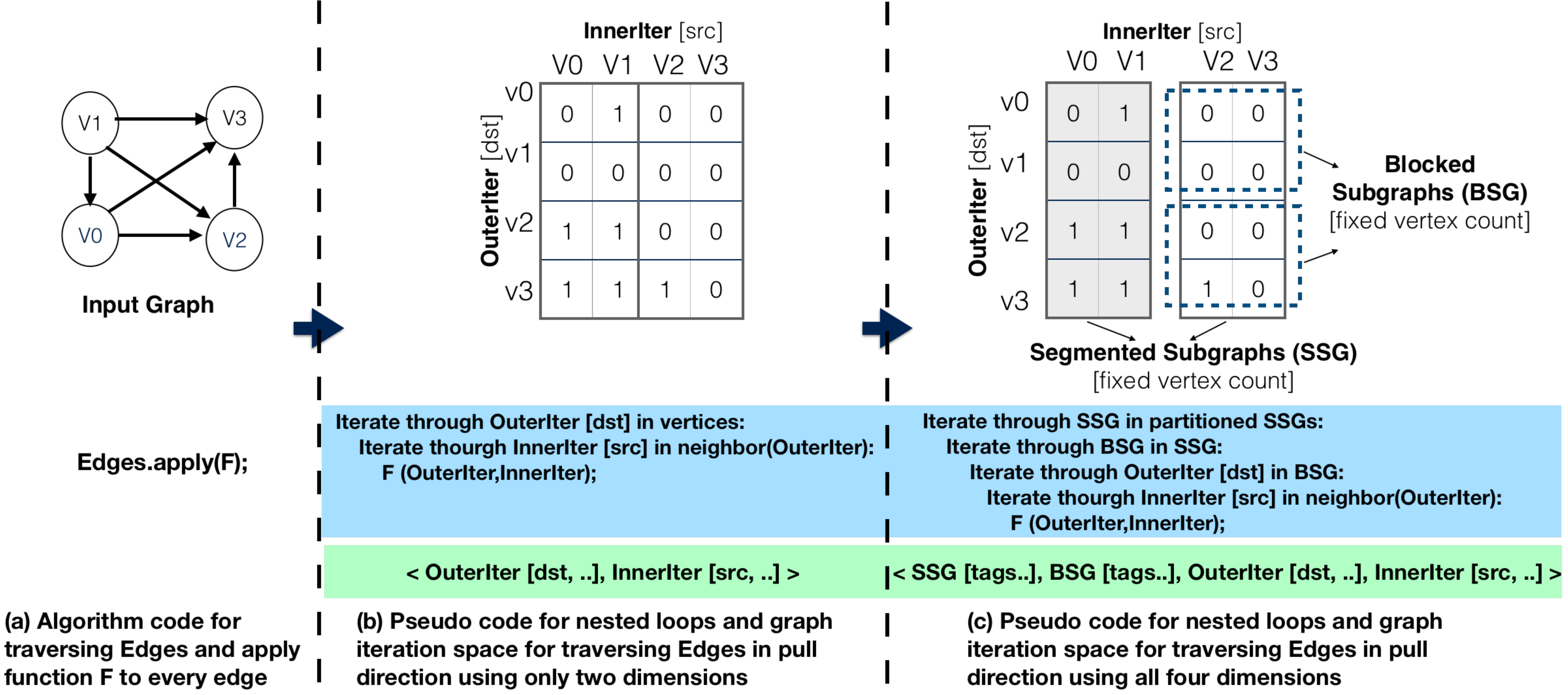}
\caption{Representing an edge traversal as nested loops and corresponding graph iteration spaces. Subfigure (c) shows the four dimensions for the graph iteration space, assuming using pull direction and the fixed vertex count partitioning strategy for both SSG and BSG dimensions.}
\label{fig:GIS_dimensions}	
\end{figure}

\begin{figure}
\begin{scriptsize}
\vspace{-4pt}
\begin{equation*}
\left\{\begin{tabular}{c|p{7.5cm}}
					   & src\_set =  filtered $src$ \textsf{vertexset} of \textbf{F}, dst\_set = filtered $dst$ \textsf{vertexset} of \textbf{F}\\
				       & $O$ $\in$ src\_set $\wedge$ $I$ $\in$ dst\_set or $O$ $\in$ dst\_set $\wedge$ $I$ $\in$ src\_set\\
					& ssg\_set = subgraphs created by segmenting the graph based on \innerIter \\
	$\langle$ $S$ [tags], $B$ [tags], $O$ [tags], $I$ [tags] $\rangle$
					& bsg\_set = subgraphs created by blocking the graph based on \outerIter \\
				       & $S$ (Segmented Subgraph ID) $\in$ ssg\_set \\
				       & $B$ (Blocked Subgraph ID) $\in$ bsg\_set \\
				       & $\langle$ $O$, $I$ $\rangle \in$ edges within the subgraph ($B$ or $S$)
				       \\
\end{tabular}\right\}
\end{equation*}
\vspace{-4pt}
\end{scriptsize}
\caption{Definition of the graph iteration space with four dimensions. $S$, $B$, $O$, and $I$ are abbreviations for \SSG, \BSG, \outerIter, and \innerIter.}
\label{fig:GIS}
\end{figure}

\subsection{\textbf{Graph Iteration Space}}

\myparagraph{Motivation} The graph iteration space is an abstract model for edge traversals
that represents the edge traversal optimizations specified by the scheduling commands in Table~\ref{table:schedule_api}. 
The model simplifies the design of the compiler 
by representing different combinations of optimizations
as multi-dimensional vectors. This representation 
 enables the compiler to easily compose together different optimizations,
  reason about validity through dependence analysis, 
  and generate nested loop traversal code.
The graph iteration space also defines the space of edge traversal 
optimizations supported by \graphit, revealing new
combinations of optimizations not explored
by prior work. 

\myparagraph{Definition} 
Let us assume that we have an operation that traverses edges and applies
a user-defined function \textbf{F} on an edgeset \textsf{Edges} as shown
in \Fig~\ref{fig:GIS_dimensions}(a).
A graph iteration space defines the set of directed edges on which 
\textbf{F} is applied and the strategy of traversing the edges.
The concept of graph iteration space is inspired by the traditional 
iteration spaces in dense nested loops~\cite{wolf1991loop, padua1986advanced}.
First, we represent the graph as an adjacency matrix, where a column in a row 
has a value of one if the column represents a neighbor of the current row (the top part of \Fig~\ref{fig:GIS_dimensions}(b)).
With this representation, we can traverse through all edges 
using dense two-level nested for loops that iterate through
 every row and every column. The traversal can be viewed 
 as a traditional 2-D iteration space. Unlike the dense iteration 
 space, the edge traversal only happens when there is an edge 
 from the source to the destination. Thus, we can eliminate unnecessary traversals
  and make the loops sparse 
by iterating only through columns with non-zero values in each row (the blue part in \Fig~\ref{fig:GIS_dimensions}(b)).
We define the row iterator variable as \textit{OuterIter}, 
and the column iterator variable as \textit{InnerIter}. The green part of
\Fig~\ref{fig:GIS_dimensions}(b) shows that a two dimensional 
graph iteration space vector is used to represent
 this two-level nested traversal.
The two-level nested for loops can be further blocked and segmented into up to 
four dimensions as shown in \Fig~\ref{fig:GIS_dimensions}(c).
The dimensions of the graph iteration space encode the nesting level of the 
edge traversal, and the tags for each dimension specify the strategy used 
to iterate through that dimension. We provide more details of the graph 
iteration space below.
\punt{Edge traversal optimizations in \graphit are encoded as a 
graph iteration space vector with tags. }
} 

\punt{

The directed edges in this set are represented
as tuples and the order among
these tuples is the lexicographical order.

\begin{figure}
\begin{scriptsize}
\vspace{-4pt}
\begin{equation}
\left\{\begin{tabular}{c|p{7cm}}
					   & $\langle outer, inner \rangle$ can be $\langle$ src, dst $\rangle$ or $\langle$ dst, src $\rangle$ \\
				       & $src$ $\in$ filtered source \textsf{vertexset} of \textbf{L}\\
	$\langle$ \SSG (S), \BSG (B), \outerIter (O), \innerIter (I) $\rangle$
					& $dst$ $\in$ filtered destionation \textsf{vertexset} of \textbf{L}\\
				       & \textit{\BSG} (Blocked Subgraph ID) identifies a subgraph created by partitioning the graph on the OuterIter dimension \\
				       & \textit{\SSG} (Segmented Subgraph ID) identifies a subgraph created by partitioning the graph based on the InnerIter dimension \\
				       & $\langle$ src, dst $\rangle \in$ edges within the immediate enclosing dimension (\BSG or \SSG)
				       \\
\end{tabular}\right\}
\end{equation}
\vspace{-4pt}
\end{scriptsize}
\caption{The definition of the graph iteration space with four dimensions}
\label{fig:GIS}
\end{figure}
}

\myparagraph{Graph Iteration Space Dimensions}
The graph iteration space in \graphit uses four
dimensions, defined in \Fig~\ref{fig:GIS} and 
illustrated in \Fig~\ref{fig:GIS_dimensions}. 
The dimensions are $\langle$ \SSG, \BSG, \outerIter, \innerIter $\rangle$ 
and are abbreviated as $\langle$ $S$, $B$, $O$, $I$ $\rangle$.
Unused dimensions are marked with $\bot$.

\outerIter($O$) and \innerIter ($I$) in \Fig~\ref{fig:GIS} are the vertex IDs of an edge (\Fig~\ref{fig:GIS_dimensions}(b)).
The ranges of $O$ and $I$ dimensions depend on the direction.
For the push direction, $O$ is in the filtered source \textsf{vertexset} (src\_set) 
and $I$ is in the filtered destination \textsf{vertexset} (dst\_set).
For the pull direction, $O$ is in the dst\_set and $I$ is in the src\_set. 
The \outerIter dimension sequentially
accesses vertices, while the \innerIter dimension has a random
access pattern due to neighbor vertex IDs not being sequential.
The edge $(O, I)$ is in
the \textsf{edgeset} of the subgraph identified by \BSG and \SSG.

The \BSG (Blocked Subgraph ID) dimension identifies a Blocked Subgraph (BSG)
in the Blocked Subgraphs Set (bsg\_set). The bsg\_set is created by partitioning the
graph by the \outerIter dimension as illustrated in the top part of \Fig~\ref{fig:GIS_dimensions}(c).
This partitioning transforms the loops that traverse the edges without
changing the graph data structure.
The graph can be partitioned with a grain size on
the number of \outerIter vertices or on the total number of edges
per BSG, depending on the schedule. This dimension controls the 
different strategies for parallelization optimizations.

\punt{This dimension controls parallel execution and load balancing optimizations.}

The \SSG (Segmented Subgraph ID) identifies a Segmented Subgraph (SSG)
 in the Segmented Subgraphs Set (ssg\_set). The 
ssg\_set is created by partitioning the graph by the \innerIter
dimension as demonstrated in the top part of \Fig~\ref{fig:GIS_dimensions}(c). 
The partitioning transforms both the graph data structure and the 
loops that traverse the edges. Details of the partitioning scheme
are described in prior work~\cite{Yunming2017}.
This dimension
controls the range of random accesses, enabling cache and
NUMA optimizations.
 The ordering of the dimensions ensures that the graph is 
segmented into SSGs before each SSG is blocked into BSGs. 
\punt{
The partitioning can be based on a fixed range of vertices in the
\innerIter dimension or a flexible range that takes into account the number of
edges depending on the schedule.  
}

\begin{figure}
\centering
\includegraphics[width=\textwidth]{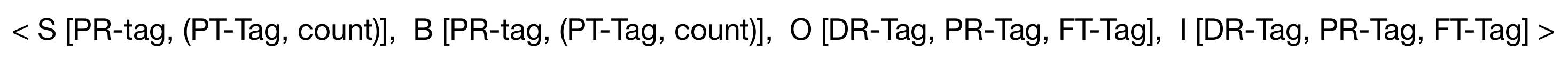}
\caption{Graph Iteration Space Tags: Direction Tags (DR-Tag), Partitioning Tags (PT-Tag),
Parallelization Tags (PR-Tag), and Filtering Tags (FT-Tag) (explained in \Fig~\ref{fig:GIS_supported}) specify direction and 
optimization strategy for each dimension, and are shown in square brackets next to each dimension.}
\label{fig:vector_tags}
\end{figure}

\begin{table}[t]
\footnotesize
  \caption{Mapping between GraphIt's scheduling language functions to the relevant dimensions and tags (highlighted in bold) of the graph iteration space.}
 \begin{tabular}{p{6.1cm}p{7.2cm}}
\textbf{Apply Scheduling Functions}  & \textbf{Graph Iteration Space Dimensions and Tags Configured} \\ \hline
 \ftt{program->configApplyDirection(label, config);} & $\langle$ $S$ [tags], $B$ [tags],  $\bm{O}$ [\textbf{direction tag, filtering tag}], $\bm{I}$ [\textbf{direction tag, filtering tag}] $\rangle$. Note, for hybrid directions (e.g. DensePull-SparsePush), two graph iteration space vectors are created, one for each direction.  \\ \hline
 \ftt{program->configApplyParallelization(label, config, [grainSize], [direction]);} &  $\langle$ $S$ [tags], $\bm{B}$ [\textbf{partitioning tag, parallelization tag}],  $O$ [tags], $I$ [tags] $\rangle$ \\ \hline
 \ftt{program->configApplyDenseVertexSet(label, \newline config, [vertexset], [direction])} & 
 	$\langle$ $S$ [tags], $B$ [tags],  $\bm{O}$ [\textbf{filtering tag}], $\bm{I}$ [\textbf{filtering tag}] $\rangle$
  \\ \hline
 \ftt{program->configApplyNumSSG(label, config, \newline numSegments, [direction]);} & 
 	$\langle$ $\bm{S}$ [\textbf{partitioning tag}], $B$ [tags],  $O$ [tags], $I$ [tags] $\rangle$
 \\ \hline
 \ftt{program->configApplyNUMA(label, config, \newline [direction]);} &  
 	 	$\langle$ $\bm{S}$ [\textbf{parallelization tag}], $B$ [tags],  $O$ [tags], $I$ [tags] $\rangle$
 \\ \hline
\noalign{\smallskip}
  \end{tabular}
  \label{table:schedule_gis_map}
\end{table}

\begin{table}
\footnotesize
\caption{The schedules applied for PageRankDelta and the generated graph iteration space vectors with tags, following the examples in \Fig~\ref{fig:prdelta_generated}. Newly added scheduling commands and the affected dimensions and tags in the graph iteration space are highlighted in bold. $\bot$ is an unused dimension. The abbreviated dimensions and tags are defined in \Fig~\ref{fig:GIS} and \Fig~\ref{fig:GIS_supported}. The keyword `Program' and the continuation
    symbol `->' are omitted. `ca' is an abbreviation for
    `configApply'. "caParallel" is short for configApplyParallelization. Note that configApplyNumSSG uses an integer
    parameter ($X$) which is dependent on the data and hardware system.
}
\begin{tabular}{p{5.9cm} p{7.1cm}}
\textbf{PageRankDelta Schedules} & \textbf{Graph Iteration Space} \\
\hline\hline

\ftt{caDirection("s1", "SparsePush");}  & $ \langle$ $\bot$, $\bot$, $O$ [$src$, SR, SA], $I$ [$dst$, SR] $\rangle $ \\ \hline
\ftt{capDirection("s1",\textbf{"DensePull-SparsePush"});} & runtime decision between two graph iteration space vectors \newline 
$\langle$ $\bm{\bot}$, $\bm{\bot}$, $\bm{O}$ [$\bm{dst}$, \textbf{SR}], $\bm{I}$ [$\bm{src}$, \textbf{SR}, \textbf{BA}] $\rangle $ and
\newline $ \langle$ $\bot$, $\bot$, $O$ [$src$, SR, SA], $I$ [$dst$, SR] $\rangle $ \\ \hline

\ftt{caDirection("s1","DensePull-SparsePush"); \newline 
     \textbf{caParallel("s1","dynamic-vertex-parallel");} }
& runtime decision between two graph iteration space vectors \newline
$\langle$ $\bot$, $\bm{B}$ [\textbf{WSP, (FVC, 1024)}], $O$ [$dst$, SR], $I$ [$src$, SR, BA] $\rangle $ and  $ \langle$ $\bot$, $\bm{B}$ [\textbf{WSP, (FVC, 1024)}], $O$ [$src$, SR, SA], $I$ [$dst$, SR] $\rangle $
 \\ \hline

\ftt{caDirection("s1","DensePull-SparsePush"); \newline 
     caParallel("s1","dynamic-vertex-parallel"); \newline
     \textbf{caDenseVertexSet("s1","src-vertexset", \newline "bitvector","DensePull");}}
& runtime decision between two graph iteration space vectors \newline 
$\langle$ $\bot$, $B$ [WSP, (FVC, 1024)], $O$ [$dst$, SR], $I$ [$src$, SR, \textbf{BV}] $\rangle $ and
\newline $ \langle$ $\bot$, $B$ [WSP, (FVC, 1024)], $O$ [$src$, SR, SA], $I$ [$dst$, SR] $\rangle $ \\ \hline

\ftt {caDirection("s1","DensePull-SparsePush"); \newline 
     caParallel("s1","dynamic-vertex-parallel"); \newline
     caDenseVertexSet("s1","src-vertexset", \newline "bitvector","DensePull");\newline
     \textbf{caNumSSG("s1","fixed-vertex-count",X,\newline  "DensePull");}}
& runtime decision between two graph iteration space vectors \newline
$\langle$ $\bm{S}$ [\textbf{SR}, (\textbf{FVC, num\_vert / $\mathbf{X}$})], $B$ [WSP, (FVC, 1024)], $O$ [$dst$, SR], $I$ [$src$, SR, BV] $\rangle $ and
\newline $\langle$ $\bot$, $B$ [WSP, (FVC, 1024)], $O$ [$src$, SR, SA], $I$ [$dst$, SR] $\rangle $ \\ \hline

\end{tabular}

\label{table:pagerankdelta_gis}
\end{table}

\myparagraph{Graph Iteration Space Tags}
\updated{
Each dimension is annotated with tags to specify the direction and 
 optimization strategies (\Fig~\ref{fig:GIS_supported} illustrates the tags in \graphit).
There are four types of tags: Direction Tags (DR-Tag), Partitioning Tags (PT-Tag),
 Parallelization Tags (PR-Tag), and Filtering Tags (FT-Tag). 
We show tags for each dimension within square brackets in
\Fig~\ref{fig:vector_tags}. 
Table~\ref{table:schedule_gis_map} shows the mapping between scheduling
language commands from Section~\ref{sec:schedule} and the corresponding
graph iteration space vector and tags.
} 
\punt{
{\footnotesize
$\langle$ \SSG [P-Tag, (PT-Tag, Count)], \BSG [P-Tag, (PT-Tag, Count)], \outerIter [P-Tag, Ftag], \innerIter [P-Tag, Ftag] $\rangle$
}
}

\updated{
Direction Tags specify whether the traversal is in push or pull direction.
In the push direction, the \outerIter is tagged as $src$ and
 \innerIter tagged as $dst$; in the pull direction, the tags  are reversed.
 Partitioning Tags specify the strategy used
for partitioning the \SSG or \BSG dimensions. For example, the default fixed vertex count (FVC) 
partitioning strategy will partition the graph based on a fixed number of \innerIter or \outerIter vertices as shown in ~\Fig~\ref{fig:GIS_dimensions}(c).
Depending on the input, this scheme may lead to an unbalanced number of edges in each SSG or BSG.
Alternatively, the edge-aware vertex count (EVC) scheme 
partitions each subgraph with a different number of \innerIter or \outerIter vertices to ensure each subgraph have similar number of edges. The EVC tag is used when the users specify the edge-aware-dynamic-vertex-parallel option with \stt{configApplyParalllelization} or the edge-aware-vertex-count option with \stt{configApplyNumSSG}.
}

\punt{
This way, each subgraph can have a similar number of edges, 
improving load balance for some algorithms. 
}

Parallelization Tags 
control whether to iterate through the 
dimension using serial (SR), static-partitioned parallel (SP), or dynamic work-stealing parallel (WSP)
execution strategies.
The PR-Tag for the \BSG dimension controls the parallelization strategy 
across different Blocked Subgraphs within a Segmented Subgraph.
Tagging the \SSG dimension to be parallel enables NUMA optimizations by executing multiple SSGs
in different sockets in parallel. If work-stealing is enabled, threads on one socket can steal unprocessed
  SSGs from another socket to improve load balance.

Filtering Tags on the \outerIter and \innerIter dimensions control the underlying data structure.
Filtering is implemented with sparse arrays (SA), dense boolean arrays (BA), or dense bitvectors (BV).
The sparse arrays contain all of the vertices that pass the filtering, while the
dense boolean arrays or the bitvectors
set the value to true or the bit to one for each vertex that passes the filtering.

\updated{
\myparagraph{Graph Iteration Spaces for PageRankDelta}
Table~\ref{table:pagerankdelta_gis} continues to use PageRankDelta 
as an example to illustrate how scheduling language commands 
generate graph iteration space vectors and tags. 
The first row shows that the SparsePush schedule maps to a graph iteration 
space vector with only two dimensions used ($\bot$ means the dimension is unused). 
The direction tags for the \outerIter and \innerIter dimensions, $src$ and $dst$, indicate that this graph iteration space 
is for the push direction. 
Going from \SparsePush to DensePull-SparsePush creates
a new graph iteration space vector for the pull direction. 
A runtime threshold on the size of the source \textsf{vertexset} 
is used to decide which vector gets executed. 
The \stt{configApplyParallelization} function configures the \BSG dimension with 
work-stealing parallelism (WSP) and uses the default 1024 grainsize. 
The fourth row demonstrates that \stt{configDenseVertexSet} sets
 the filtering tag for the innerIter dimension to 
bitvector (BV) in the graph iteration space vector for the pull direction. 
Finally, \stt{configNumSSG} sets up the \SSG dimension to partition the 
graph for cache locality. In the fixed-vertex-count configuration (FVC),
 the \innerIter range for each SSG is computed by dividing 
the total number of vertices by the number of SSGs specified with $X$. 

\punt{ 
\julian{it is strange that num\_vert appears in the graph iteration space because we do not know the total number of vertices until runtime. what actually gets passed to the compiler?}
}
  
\punt{
Optimizations on different dimensions of the graph iteration space, such as
cache-aware graph segmenting (\SSG) and direction optimization (\outerIter and \innerIter), can be combined together.
Optimizations working on different tags for the same dimension, such as fixed vertex count partitioning (FVC) and work-stealing parallelism (WSP),  can also be composed.
However, optimizations working on the same tags of the same dimension, such as working-stealing parallelism (WSP) and static parallelism (SP), cannot be composed.

Programmers can form composite graph iteration spaces.
 The hybrid directions, such as DensePull-SparsePush,
 can be represented as two separate graph iteration spaces.
A runtime threshold on the size of the source \textsf{vertexset}
can be used in this case to decide which version gets executed.
Table~\ref{table:vector_tags_examples} shows a few examples of optimizations
expressed as graph iteration spaces and tags.
}

\myparagraph{Generalizing Graph Iteration Spaces} The graph iteration space concept
can be generalized to expand the space of supported optimizations. 
In \graphit, we restrict the graph iteration space to four dimensions with fixed 
partitioning schemes. 
Adding more dimensions and/or removing constraints on how the dimensions are partitioned 
can potentially represent additional optimizations.

\subsection{Vertex Data Layout and Program Structure Optimizations Representation} Since the vertex data are stored as abstract vectors,
 they can be implemented as an array of structs or struct of arrays. We use vector tags
 to tag each vertex data vector as Array of Structs (AoS) or a separate array in the implicit global struct (SoA). These tags can be configured with the \stt{fuseFields} scheduling function.
 \punt{\julian{tie this back to section 4}}
Program structure optimizations update the structure of the loops and edgeset \apply{} operators. We use the scoped labels (described in Section~\ref{subsec:structural}), which are specified in the scheduling language with \stt{fuseForLoop} and \stt{fuseApplyFunctions}, to represent the optimizations. 
\punt{\julian{we did not describe what IR nodes are}}

}

\section{Compiler Implementation}
\label{sec:compiler}

\punt{
The organization of compiler passes is
shown in Figure~\ref{fig:compilerorganisation}.
}

This section describes the \graphit compiler, which generates optimized C++ code from
an algorithm and a schedule. We also built an autotuner on top of the compiler to 
automatically find high-performance schedules. 

\punt{
Fig.~\ref{fig:codegen} illustrates the main steps
 to generate code for the \stt{main} function in a \graphit program,
the user-defined \stt{apply} functions, and edge traversals.
}


\begin{figure}[t]
\centering
\includegraphics[width=0.7\textwidth]{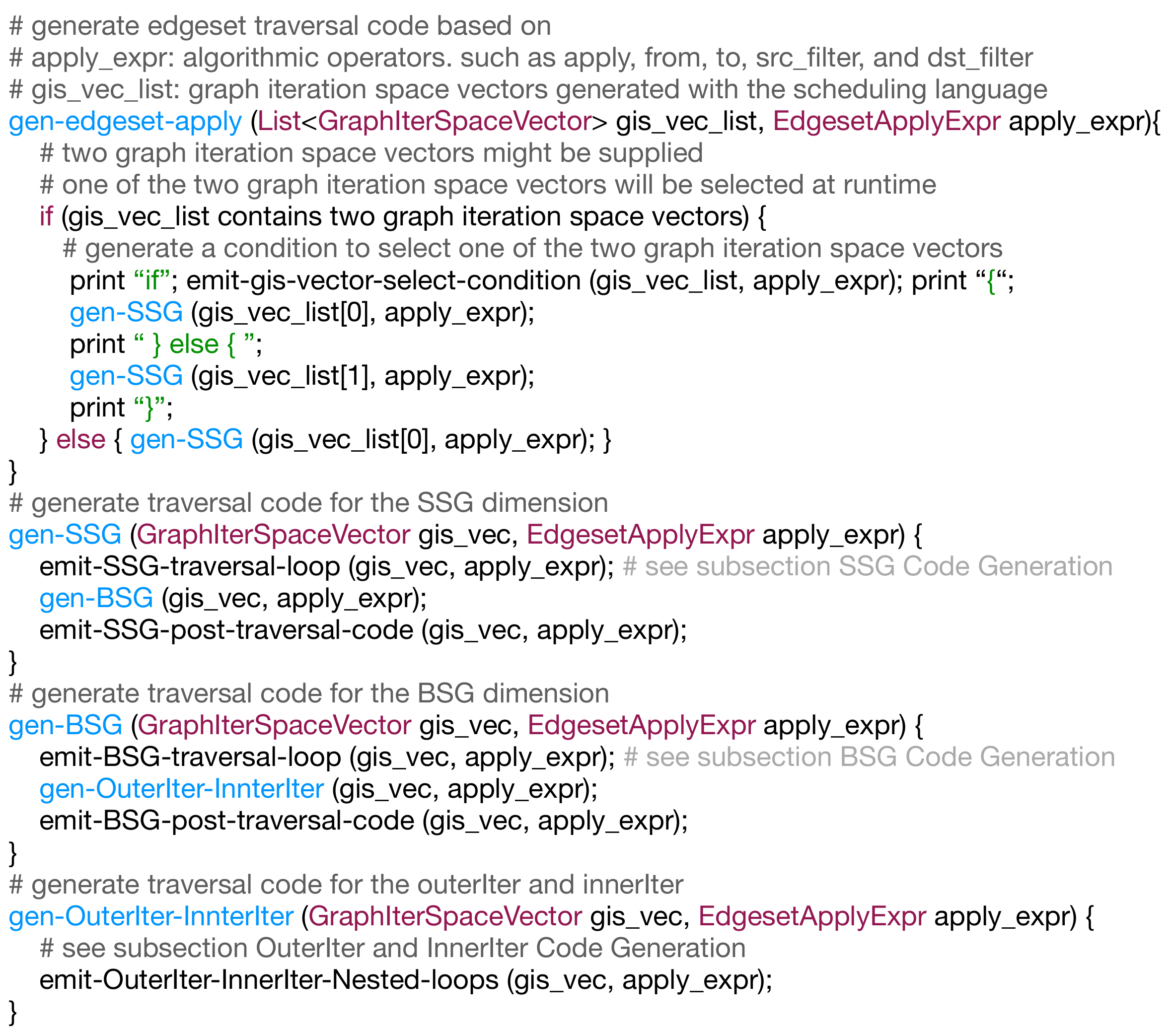}
\caption{Code generation algorithm for the graph iteration space.}
\label{fig:gis_code_gen}
\end{figure}

\subsection{Code Generation for Graph Iteration Space}
\updated{
We first show the high-level code generation 
algorithm for the graph iteration space in \Fig~\ref{fig:gis_code_gen}.
To deal with hybrid traversal modes that have two graph iteration space vectors,
 such as DensePull-SparsePush, 
\stt{gen-edgeset-apply} 
generates two implementations of edge traversal logic with additional
logic to choose an implementation
based on the sum of the out-degrees of active vertices (the "if", "else", and \stt{emit-gis-vector-select-condition} shown in ~\Fig~\ref{fig:gis_code_gen})
 as described in Section~\ref{sec:tradeoff}. The functions 
 \stt{gen-SSG}, \stt{gen-BSG}, and \stt{gen-OuterIter-InnerIter} 
generate nested traversal loops for the different 
 graph iteration space dimensions. Below, we
 provide more details on the code generation functions and the mechanisms
 to ensure the validity of the optimizations. 
}

 \myparagraph{\outerIter and \innerIter Code Generation} We demonstrate how to
 generate traversal code for the \outerIter and \innerIter dimensions
 using a simple example with the 
 \SparsePush configuration shown in Fig.~\ref{fig:sparsePushConfig}
 (graph iteration space and tags: $ \langle$ $\bot$, $\bot$, $O$ [$src$, SR, SA], $I$ [$dst$, SR, BA] $\rangle$; abbreviations and sets are defined in  Fig.~\ref{fig:GIS} and Fig.~\ref{fig:GIS_supported}).

 \begin{wrapfigure}[4]{r}{0.6\textwidth}
\begin{lstlisting}[language=graphit]
	#s1# edges.from(Frontier).dstFilter(dstFunc).apply(applyFunc)
	schedule:
	program->configApplyDirection("s1","SparsePush");
\end{lstlisting}
\caption{SparsePush configuration.}\label{fig:sparsePushConfig}
\end{wrapfigure}

For the push direction, \outerIter is $src$ and \innerIter is $dst$. Since the source (\outerIter)
filtering is tagged as Sparse Array (SA),
the outer loop iterates over the source \textsf{vertexset} (Frontier). The \stt{dst} (\innerIter) filtering
uses the user-defined boolean function \stt{dstFunc}. 
The generated code is shown in Fig.~\ref{fig:sparsePush}.

\punt{$: $O$ $\in$ dst\_set, I $\in$ src\_set $}
We show code generated for a DensePull traversal mode  ($ \langle$ $\bot$, $\bot$, $O$ [$dst$, SR, BA], $I$ [src, SR, BA] $\rangle$ ) in Fig.~\ref{fig:densePull}.
The \outerIter is now $dst$ and the \innerIter is $src$.
The user-defined function \stt{applyFunc}
is applied to every edge as before.
The \textsf{vertexset}s are automatically converted from the sparse array of vertices
(\stt{vert\_array} shown in the ~\SparsePush~ example above) to a
boolean map (\stt{bool\_map} in the~\DensePull~example).
Filtering on
destination vertices (\dstFilter) is attached as an \stt{if}
statement next to the $dst$ iterator (\outerIter).

\begin{figure}
   \begin{minipage}{.48\textwidth}
\begin{lstlisting} [language = c++]
	for (int i = 0; i < Frontier.size(); i++){
		NodeID src = Frontier.vert_array[i];
		for (NodeID dst : G.getOutNghs(src)){
			if (dstFunc(dst)){
				applyFunc(src, dst); }}}

\end{lstlisting}
\captionof{figure}{Generated SparsePush code.}\label{fig:sparsePush}
   \end{minipage}
\begin{minipage}{.48\textwidth}

\begin{lstlisting} [language = c++]
	for (NodeID dst = 0; dst < num_verts; dst++){
		if (dstFunc){
			for (NodeID src : G.getInNghs(dst)){
				if (Frontier.bool_map(src)){
					applyFunc(src, dst); }}}}
\end{lstlisting}
\captionof{figure}{Generated DensePull code.}\label{fig:densePull}
\end{minipage}
\end{figure}



\myparagraph{Blocked Subgraph (BSG) Code Generation}
 The BSG dimension in the graph iteration space is created by partitioning the
\outerIter dimension.
\graphit uses the partitioning tag for this dimension to control
the granularity and blocking
strategy for load balancing,
and the parallelization tag to control
the mode of parallelization.
\Fig~\ref{fig:bsg} shows an example of the generated code,
 assuming \outerIter represents $src$.

 \begin{wrapfigure}{r}{0.7\textwidth}
\begin{lstlisting} [language = c++]
parallel_for (int BSG_ID = 0; BSG_ID < g.num_chunks; BSG_ID++){
	for (NodeID src = g.chunk_start[BSG_ID]; src < g.chunk_end[BSG_ID]; src++)
		for (NodeID dst : G.getOutNghs(src))
			applyFunc(src,dst);}
\end{lstlisting}
\caption{Generated BSG code.}\label{fig:bsg}
\end{wrapfigure}

If the edge-aware vertex count (EVC) partitioning tag is used,
the compiler generates chunks with approximately the number of edges specified
by the schedule. For the fixed vertex count (FVC) partitioning tag,
the compiler uses the built-in grain size in OpenMP.
For parallelization tags static parallelism (SP) and dynamic work-stealing parallelism (WSP), we
simply use the OpenMP pragmas
to implement \stt{parallel\_for} 
(\stt{pragma omp for parallel schedule (static)}
and \stt{schedule (dynamic)}).

\myparagraph{Segmented Subgraph (SSG) Code Generation}
Using the SSG dimension requires adding a loop outside of
 the existing traversals and changing the data layout of the graph.
 \graphit generates code in the main function to create 
 the SSGs by partitioning the graph by \innerIter. 
 This partitioning can use a fixed range of vertices (FVC) in the \innerIter
 or a flexible range of vertices that takes into account the number of edges in each
 SSG (EVC) with an edge grain size. The random memory access range in each SSG is restricted to
 improve locality.
Fig.~\ref{fig:ssg} shows edge traversal code that
 uses both SSG and BSG dimensions
 ($ \langle$ $S$ [SR], $B$ [SR], $O$ [$dst$, SR], $I$ [$src$, SR] $\rangle $). The segmented subgraphs are stored in \stt{g.SSG\textunderscore}list.

\punt{
$$O$ $\in$ dst\_set, $I$ $\in$ src\_set, B $\in$ bsg\_set, $S$ $\in$ ssg\_set  $
 }

  \begin{wrapfigure}{r}{0.75\textwidth}

\begin{lstlisting} [language = c++]
for (int SSG_ID = 0; SSG_ID < num_SSG; SSG_ID++){
	sg = g.SSG_list[SSG_ID];
	for (int BSG_ID = 0; BSG_ID < sg.num_chunks; BSG_ID++){
		for (NodeID dst = sg.chunk_start[BSG_ID]; dst < sg.chunk_end[BSG_ID]; dst++)
			for (NodeID src : G.getInNghs(dst))
				applyFunc(src,dst);}}
\end{lstlisting}
\caption{Generated SSG and BSG code.}\label{fig:ssg}
\end{wrapfigure}

The cache optimization processes one SSG at a time (SR), but processes the
BSGs within the SSG in parallel.
The programmer can enable NUMA optimizations by
specifying the parallelization tag for SSG as static parallel (SP);
the compiler then assigns different SSGs to be executed on
different sockets. \graphit implements this assignment
using \stt{numa\textunderscore alloc}
in the main function to first allocate SSGs on different
sockets, and then uses the \stt{proc\textunderscore bind} API in OpenMP to assign threads
to process each socket-local subgraph. If work-stealing parallelism (WSP)
is enabled for SSGs, then a socket can steal an SSG allocated on
another socket if no work remains on the current socket.

In some cases, we need to use NUMA-local buffers to store intermediate
results from each SSG. The compiler generates code for allocating
NUMA-local buffers and changes the data references from updating
global vertex data vectors to the NUMA-local buffers.
The compiler also generates code for a merge phase that
merges NUMA-local buffers to update the global data vectors.


\myparagraph{Validity of Optimizations} \graphit ensures
the validity of single edge traversal optimizations
by imposing a set of restrictions on the \graphit
language and using dependence analysis to insert
 appropriate atomic synchronization instructions.

\punt{The language restrictions in \graphit are imposed
on vertex data vector accesses across \srcFilter{}, \dstFilter{}, and
\apply{} functions for any given \textsf{edgeset}.}

We enforce some restrictions on read-write accesses and reduction 
operators for vertex data vectors across user-defined functions used in 
 \srcFilter{}, \dstFilter{}, and edgeset 
\apply{} functions for a given \textsf{edgeset} traversal operation. 
Each vertex data vector must have only one of the following properties: read-only, write-only, or reduction. 
Additionally, reductions are commutative and associative.
With these two restrictions, transformations do not need to preserve
 read-after-write dependences, and
transformations remain valid independent
of edge traversal order. Therefore, a transformed program is valid as
long as each filtered edge is processed exactly once.
\graphit does provide \stt{asyncMax} and \stt{asyncMin} reduction operators for cases where
there can be some read and write dependences. The programmer is responsible for
ensuring that reordering traversals will not affect the final outcome of the
program when using \stt{asyncMax} and \stt{asyncMin}. These operators are useful for
applications such as connected components and single-source shortest paths
as the vertex data values will eventually converge.

\punt{
\graphit can also easily be extended to support multiple
read-write property by leveraging fine-grained locking on sources and
destinations. However, currently, we focus on high-performance atomic
synchronization on individual vector accesses.
}

To ensure that each filtered edge is processed exactly once,
we insert synchronization code to vertex data vector updates
by leveraging dependence analysis theory
from dense loop iteration spaces~\cite{maydan1991efficient,li1989data}.
Dependence analysis is well-suited for \graphit
because the language prevents aliasing and each vertex data vector
represents a separate data structure. Additionally, the goal of the
analysis is not to automatically parallelize the loop with a
correctness guarantee, but the much easier task of determining whether
synchronization code is necessary for a given parallelization
scheme. Accomplishing this task does not require a precise distance
vector.

Below we show a code snippet of PageRankDelta with the SparsePush configuration (
$ \langle$ $\bot$, $\bot$, $O$ [$src$, SR, SA], $I$ [$dst$, SR] $\rangle$), the distance vector, and read-write properties of the vectors.
\punt{$: $O$ $\in$ src\_set, $I$ $\in$ dst\_set $}

\begin{minipage}{0.47\textwidth}
\vspace{3mm}
\begin{lstlisting} [language = c++]
for (int i = 0; i < Frontier.size(); i++){
  NodeID src = Frontier.vert_array[i];
  for (NodeID dst : G.getOutNghs(src)){
      DeltaSum[dst] += Delta[src]/OutDegree[src];
      }}}
\end{lstlisting}
\end{minipage}
\begin{minipage}{0.49\textwidth}
\vspace{-1mm}
\begin{footnotesize}
\begin{tabular}{l c c}
	{Vector Dependence} & {Distance Vector} & {Read-Write} \\ \hline
	DeltaSum & $\langle *, 0 \rangle$ & reduction \\
	Delta & $\langle 0, 0 \rangle$ & read-only \\
	OutDegree & $\langle 0, 0 \rangle$ & read-only
\end{tabular}
\end{footnotesize}
\end{minipage}

\punt{ 
\julian{where are the annotations in the code? how would the annotations work if the vertex data vectors have different read-write properties in different parts of the algorithm?}
}
The compiler builds a dependence vector for the \outerIter and \innerIter dimensions
based on their direction tags.
We see that DeltaSum has a dependence with
the reduction operator (it is both read from and written to).
Different source nodes
can update the same $dst$, and so we assign $*$ to the
first element of the distance vector to denote that there is a
dependence on an unknown iteration of $src$,
which maps to \outerIter based on the direction tags.
Given a $src$, we know that the
$dst$'s are all different, and thus, there is no data dependence on
the second iterator and we assign the second value of the distance
vector as $0$. Since Delta and OutDegree are both read-only, they have
the distance vector $\langle 0,0 \rangle$ with no dependence across
different iterations.  Given that DeltaSum's distance vector's first
element is $*$, the compiler knows that synchronization must be
provided when parallelizing the \outerIter (outer loop).  If
 only the \innerIter (inner loop) is parallelized, then no synchronization is needed. 

\punt{$: $O$ $\in$ dst\_set, $I$ $\in$ src\_set $}
A similar analysis works on a \DensePull ($ \langle$ $\bot$, $\bot$, $O$ [$dst$, SR, BA], $I$ [$src$, SR] $\rangle$) PageRankDelta.
The code snippet and distance vectors are shown below.
The first element in the distance vector for DeltaSum is $0$
given that there is no dependence among different destination
vertices and \outerIter represents $dst$.
However, the value is $*$ on the second element because
different sources will update the same destination. Parallelizing
\outerIter in this case does not require any synchronization.

\begin{minipage}{0.50\textwidth}
\vspace{3mm}
\begin{lstlisting} [language = c++]
for (NodeID dst = 0; dst < num_verts; dst++) {
  for (NodeID src : G.getInNghs(dst)){
      if (Frontier.bool_map(src))
         	DeltaSum[dst] += Delta[src]/OutDegree[src];
         	}}}
\end{lstlisting}
\end{minipage}
\begin{minipage}{0.49\textwidth}
\vspace{-1mm}
\begin{footnotesize}
\begin{tabular}{l c c}
	{Vector Dependence} & {Distance Vector} & {Read Write} \\ \hline
	DeltaSum & $\langle 0, * \rangle$ & reduction \\
	Delta & $\langle 0, 0 \rangle$ & read-only \\
	OutDegree & $\langle 0, 0 \rangle$ & read-only
\end{tabular}
\end{footnotesize}
\end{minipage}

Since BSG is partitioned by \outerIter, parallelizing the \BSG dimension
would have the same effect as parallelizing \outerIter. Similarly,
parallelizing \SSG has the same effect as parallelizing \innerIter
given SSG is partitioned by  \innerIter.

When applying NUMA optimizations to the second code snippet with the DensePull direction
(parallelizing both the SSG and BSG dimensions),
we have a dependence vector of $\langle *, * \rangle$ for DeltaSum.
In this case, \graphit writes the updates to DeltaSum[$dst$] to a socket-local
buffer first and later merges buffers from all sockets to provide
synchronization. 

\punt{This approach is more efficient then inserting atomic
instructions since we know that the access pattern for \outerIter is sequential.
\julian{how does sequential access pattern make this more efficient?}
\julian{socket-local buffers slows convergence for CC and SSSP so it is not necessary faster overall}}
  
\punt{
There is one exception with the following pattern. \graphit recognizes that this if statement can be replace with an atomic compare-and-swap operation.
\begin{lstlisting} [language = c++]
for (int i = 0; i < frontier.size(); i++){
  NodeID src = frontier.vert_array[i];
  for (NodeID dst : graph.out_neigh(src)){
      if (vector[dst] == old_val) vector[dst] == new_val; }}
\end{lstlisting}
}

For the hybrid traversal configurations, \graphit  generates two versions of the user-defined \apply{} function since the synchronization requirements for the push and pull directions are different. Each version will be used in the corresponding traversal mode.


\punt{

With the transformed IR, \graphit maps schedule configurations to
program elements that are non-existent in the original program by
taking advantage of the label scopes and name nodes described in
Section~\ref{sec:schedule}.  When performing lowering based on schedules,
the compiler first dynamically computes the scope of the labels for each program
element while traversing the modified IR and stops when there is a
match between the current scoped label and the label specified in the
scheduling command.

\st{Ensuring the validity of the multiple edge traversal optimizations is entirely up
to the programmer as the compiler current do not perform dependence analysis
across different edge traversals.}
}

\begin{wrapfigure}{r}{0.35\textwidth}
\begin{lstlisting} [language=graphit]
func vertexset_apply_f(v:Vertex)
	parent[v] = -1;
end
func main()
	vertices.apply(vertexset_apply_f);
end
\end{lstlisting}
\caption{\textsf{Vertexset} \apply{} code.}\label{fig:vertexcode}
\end{wrapfigure}

\subsection{Code Generation for Vertex Data Layout Optimizations}
To generate code with different
physical data layouts for the vertex data (array of structs or struct of arrays),
\graphit generates declaration and initialization code in the main function
and updates references to vertex data in the other functions.
The compiler first transforms assignments on vertex data vectors
into \textsf{vertexset} \apply{} operations that set the values of the data vectors.
If the programmer specifies the \stt{fuseField} command,
 \graphit generates a new struct type, an array of structs declaration, and changes 
 the references to the vertex data vectors in the functions to access fields of the struct instead of separate arrays. 
 For example,
 the assignment statement for the parent vector \stt{ parent : vector \{Vertex\}(int) = -1;} is implemented
by first declaring an \stt{apply} function \stt{vertexset\_apply\_f} and another
\stt{vertices.apply} statement in the main function that uses \stt{vertexset\_apply\_f}
as shown in Fig.~\ref{fig:vertexcode}.
 The vector access expression in the \stt{apply} function will then be lowered from \stt{parent[v]}
 to \stt{fused\_struct[v].parent} (Line 2 of Fig.~\ref{fig:vertexcode}). The correctness of the program is not impacted by the vertex data layout optimization as it
does not affect the execution ordering of the program.

\updated{
\subsection {Code Generation for Program Structure Optimizations}
Traditional compilers with a fixed number and order of optimization
passes are ill-suited for program structure optimizations, 
such as kernel fusion. The \graphit compiler introduces
a new schedule-driven optimization pass orchestration design that
allows users to add more optimization passes and dictate the order of
the added optimizations with the label-based scheduling language
described in Section~\ref{sec:schedule}. Users can perform fine-grained
loop fusion, loop splitting, and fusion of \apply{} functions on
loops and functions specified with statement labels and scheduling commands.
These optimizations are implemented as customized optimization passes
on the intermediate representation of the program. \graphit implements these
program structure transformation schedules by adding new optimization passes 
that transform the intermediate representation.

\subsection {Autotuning \graphit Schedules}
Finding the right set of schedules can be challenging for non-experts.
\graphit can have up to $10^5$ valid schedules with each run taking
more than 30 seconds for our set of applications and input graphs.
Exhaustive searches would require weeks of time.  As a result, we use
OpenTuner~\cite{ansel:pact:2014} to build an autotuner on top of
\graphit that leverages stochastic search techniques (e.g., AUC
bandit, greedy mutation, differential mutation, and hill climbing) to
find high-performance schedules within a reasonable amount of
time.

\myparagraph{Search space}
We limit the tuning to a single edgeset \apply{} operation identified 
by the user. 
The autotuner will try different configurations for the 
direction of the traversal (\stt{configApplyDirection}), 
the parallelization scheme (\stt{configApplyParallelization}), 
the data layout for the dense vertexset (\stt{configApplyDenseVertexSet}),
the partitioning strategy of the graph 
(\stt{configApplyNumSSG}), and the NUMA execution policy (\stt{configApplyNUMA}).

Not all generated schedules are valid because schedules 
have dependencies among them. For example, \stt{configApplyNumSSG}, which 
takes a direction parameter, 
 is only valid if the direction specified 
 is also set by \stt{configApplyDirection}. Instead of reporting an 
 invalid schedule as error, \graphit's autotuner 
 ignores invalid schedules to smooth the search space. For 
 example, the \stt{configApplyNumSSG} configuration is ignored if 
 the direction specified is not valid.  

} 

\punt{
\subsection{Other Optimizations}
Many real-world programs that extract statistics features from graphs~\cite{leskovec2016snap}
 contain multiple edge traversals
sharing the same edge traversal logic. Graph iteration space vectors and tags can be fused
together to further improve the performance of these programs.
In order to fuse different edge traversals,
we designed operations on the intermediate representations that can perform
index splitting and fusion on the loops containing the edge traversal
code. \graphit can also compose multiple edge traversal optimizations
with single edge traversal and vertex data layout optimizations
to significantly reduce the number of random
 memory accesses. We include the details of the optimization in the Appendix~\yunming{add references here}.
 }

\section{Evaluation}
\label{sec:eval}

In this section, we compare \graphit's performance to state-of-the-art
frameworks and DSLs on graphs of various sizes and structures. We also
analyze performance tradeoffs among different \graphit schedules.
We use a dual socket system with Intel Xeon E5-2695 v3 CPUs with 12
cores each for a total of 24 cores and 48 hyper-threads. The system
has 128GB of DDR3-1600 memory and 30 MB last level cache on each
socket, and runs with Transparent Huge Pages (THP) enabled.

\begin{wraptable}[10]{R}{0.55\textwidth}
  \center \tabcolsep 5pt
    \caption{Graphs used for experiments. 
    LiveJournal and Friendster are undirected, and the number of edges counts each edge in both directions. All
    other graphs are directed.}
  \scriptsize
  \begin{tabular}{l|r|r}
    Dataset & Num. Vertices &  Num. Edges \\ \hline
    \textit{LiveJournal} (LJ)~\cite{davis11acm-florida-sparse} & 5 M & 69 M  \\ \hline
    \textit{Twitter} (TW)~\cite{kwak10www-twitter} & 41 M & 1469 M \\ \hline
    \textit{WebGraph} (WB) ~\cite{sd-graph} & 101 M & 2043 M \\ \hline
    \textit{USAroad} (RD)~\cite{road-graph} & 24 M & 58 M \\ \hline
    \textit{Friendster} (FT) ~\cite{snapnets} & 65.6 M & 3.6 B \\ \hline
    \textit{Netflix} (NX)~\cite{Bennett07thenetflix} & 0.5 M & 198 M  \\ \hline
    \textit{Netflix2x} (NX2)~\cite{LiNetflix} & 1 M & 792 M \\ \hline
  \end{tabular}
  \label{table:datasets}
\end{wraptable}

\myparagraph{Data Sets} Table~\ref{table:datasets} lists our input
datasets and their sizes. LiveJournal, Twitter, and Friendster are
three social network graphs. Friendster is special because its number
of edges does not fit into a 32-bit signed integer. We use WebGraph from the
2012 common crawl. USAroad is a mesh network with small and
undeviating degrees. The Netflix rating dataset and its synthesized
expansion (Netflix2x) are used to evaluate Collaborative Filtering.

\myparagraph{Algorithms} We try to use the same algorithms across different
frameworks to study the impact of performance optimizations.  Our
evaluation is done on seven algorithms: PageRank (PR), Breadth-First
Search (BFS), Connected Components (CC) with synchronous label
propagation, Single Source Shortest Paths (SSSP) with frontier based
Bellman-Ford algorithm, Collaborative Filtering (CF), Betweenness Centrality (BC), and
PageRankDelta (PRDelta). For Galois, we used the asynchronous algorithm 
    for BFS, and the Ligra algorithm for SSSP.

\punt{ which includes PageRank (PR), Breadth-First Search (BFS),
  Connected Components (CC), Single Source Shortest Paths (SSSP),
  PageRankDelta (PRDelta), and Collaborative Filtering (CF). These
  algorithms expose diverse performance characteristics. PR and CF
  operate on the entire graph. BFS and SSSP are traversal algorithms
  (we use an active set-based version of Bellman-Ford for SSSP). Our
  implementation of CC uses label propagation, which initially
  operates on the whole graph and in subsequent iterations only
  performs computation on the vertices whose data changed in the
  previous iteration.  }

\myparagraph{Existing Frameworks} We compare GraphIt's performance to
six state-of-the-art in-memory graph processing systems: Ligra,
GraphMat, Green-Marl, Galois, Gemini, and Grazelle. Ligra has fast
implementations of BFS and SSSP~\cite{shun13ppopp-ligra}. Among prior
work, GraphMat has the fastest shared-memory implementation of
CF~\cite{sundaram15vldb-graphmat}. Green-Marl is one of the fastest
DSLs for the algorithms we evaluate~\cite{Hong12asplos}. Galois
(v2.2.1) has an efficient asynchronous engine that works well on road
graphs ~\cite{nguyen13sosp-galois}. Gemini is a distributed graph
processing system with notable shared-machine
performance~\cite{Zhu16gemni}.  Compared to existing frameworks,
Grazelle has the fastest PR and CC using edge list vectorization,
inner loop parallelism, and NUMA optimizations~\cite{Grossman2018}.


\subsection{Comparisons with State-of-the-Art Frameworks}

\label{subsec:compare_with_others}
Table~\ref{table:eval-perf-table} shows the execution time of \graphit
and other systems. 
The best performing schedules for \graphit are shown
in Table~\ref{table:eval-schedules}. Table~\ref{table:LOC} shows the
 line counts of four graph algorithms for each framework. 
 \graphit often uses significantly fewer lines of code compared to the other frameworks. Unlike \graphit, other frameworks with direction optimizations require programmers to provide many low-level implementation details as discussed in Section~\ref{sec:algo}.
\graphit outperforms the next fastest of the six state-of-the-art shared-memory
frameworks 
on \numFastExp out of \numExp experiments by up to 4.8$\times$, and is
 never more than 43\% slower than the fastest framework on the other experiments. 

 \begin{table}[t]\scriptsize
\caption{Running time (seconds) of GraphIt and state-of-the-art frameworks. The fastest results are bolded. The missing numbers correspond to a framework not supporting an algorithm and/or not successfully running on an input graph. Galois' Betweenness Centrality (BC) uses an asynchronous algorithm, while other frameworks use a synchronous one. We ran PageRank (PR) for 20 iterations, PageRankDelta (PRDelta) for 10 iterations, and Collaborative Filtering (CF) for 10 iterations. Breadth-First Search (BFS), Single Source Shortest Paths (SSSP), and Betweenness Centrality (BC) times are averaged over 10 starting points.}
\tabcolsep 1.5pt
\begin{tabular}{l|rrrrr|rrrrr|rrrrr|rr}
Algorithm & \multicolumn{5}{c|}{ PR } & \multicolumn{5}{c|}{ BFS } & \multicolumn{5}{c}{ CC } & \multicolumn{2}{c}{ CF } \\ \hline
Graph & LJ & TW & WB & RD & FT & LJ & TW & WB & RD & FT & LJ & TW & WB & RD & FT & NX & NX2\\ \hline
GraphIt & \textbf{0.342} & \textbf{8.707} & \textbf{16.393} & 0.909 & \textbf{32.571} & 0.035 & \textbf{0.298} & \textbf{0.645} & \textbf{0.216} & \textbf{0.490} & 0.068 & \textbf{0.890} & \textbf{1.960} & 17.100 & \textbf{2.630} & \textbf{1.286} & \textbf{4.588} \\
Ligra & 1.190 & 49.000 & 68.100 & 1.990 & 201.000 & \textbf{0.027} & 0.336 & 0.915 & 1.041 & 0.677 & \textbf{0.061} & 2.780 & 5.810 & 25.900 & 13.000 & 5.350 & 25.500\\
GraphMat & 0.560 & 20.400 & 35.000 & 1.190 & & 0.100 & 2.800 & 4.800 & 1.960 & & 0.365 & 9.8 & 17.9 & 84.5 & & 5.010 & 21.600 \\
Green-Marl & 0.516 & 21.039 & 42.482 & 0.931 & & 0.049 & 1.798 & 1.830 & 0.529 & & 0.187 & 5.142 & 11.676 & 107.933 & \\
Galois & 2.788 & 30.751 & 46.270 & 9.607 & 117.468 & 0.038 & 1.339 & 1.183 & 0.220 & 3.440 & 0.125 & 5.055 & 15.823 & 12.658 & 18.541 \\
Gemini & 0.430 & 10.980 & 16.440 & 1.100 & 44.600 & 0.060 & 0.490 & 0.980 & 10.550 & 0.730 & 0.150 & 3.850 & 9.660 & 85.000 & 13.772 \\
Grazelle & 0.368 & 15.700 & 20.650 & \textbf{0.740} & 54.360 & 0.052 & 0.348 & 0.828 & 1.788 & 0.512 & 0.084 & 1.730 & 3.208 & \textbf{12.200} & 5.880 \\
& & & & & & & & & & & & & & & \\
Algorithm & \multicolumn{5}{c|}{ SSSP } & \multicolumn{5}{c|}{ PRDelta } & \multicolumn{5}{c}{ BC }  & & \\ \hline
Graph & LJ & TW & WB & RD & FT & LJ & TW & WB & RD & FT &  LJ & TW & WB & RD & FT \\ \hline
GraphIt & 0.055 & \textbf{1.349} & \textbf{1.680} & \textbf{0.285} & \textbf{4.302} & \textbf{0.183} & \textbf{4.720} & \textbf{7.143} & \textbf{0.494} & \textbf{12.576} & 0.102 & 1.550 & 2.500 & \textbf{0.650} & \textbf{3.750} \\
Ligra & \textbf{0.051} & 1.554 & 1.895 & 1.301 & 11.933 & 0.239 & 9.190 & 19.300 & 0.691 & 40.800 & 0.087 & 1.931 & 3.619 & 2.530 & 6.160 \\
GraphMat & 0.095 & 2.200 & 5.000 & 43.000 & & & & & & &  &  & & & \\
Green-Marl & 0.093 & 1.922 & 4.265 & 93.495 & & & & &  & & \textbf{0.082} & 3.600 & 6.400 & 29.050 & \\
Galois & 0.091 & 1.941 & 2.290 & 0.926 & 4.643 & & & & & & 0.237 & 3.398 & 4.289 & 0.806 & 9.897 \\
Gemini & 0.080 & 1.360 & 2.800 & 7.420 & 6.147 & & & & &  & 0.149 & \textbf{1.358} & \textbf{2.299} & 31.055 & 3.849 \\ \hline
\end{tabular}
\label{table:eval-perf-table}
\end{table}

\begin{table}[t]
  \centering
    \caption{Schedules that GraphIt uses for all applications on
    different graphs. The schedules assume that the edgeset \apply{} operator is labeled with s1.
     The keyword 'Program' and the continuation
    symbol '->' are omitted. 'ca' is the abbreviation for
    'configApply'. Note that configApplyNumSSG uses an integer
    parameter ($X$) which is dependent on the graph size and the cache
    size of a system. BC has two edgeset \apply{} operators, denoted with s1 and s2.  }
  \label{table:eval-schedules}
  \includegraphics[width=1\columnwidth]{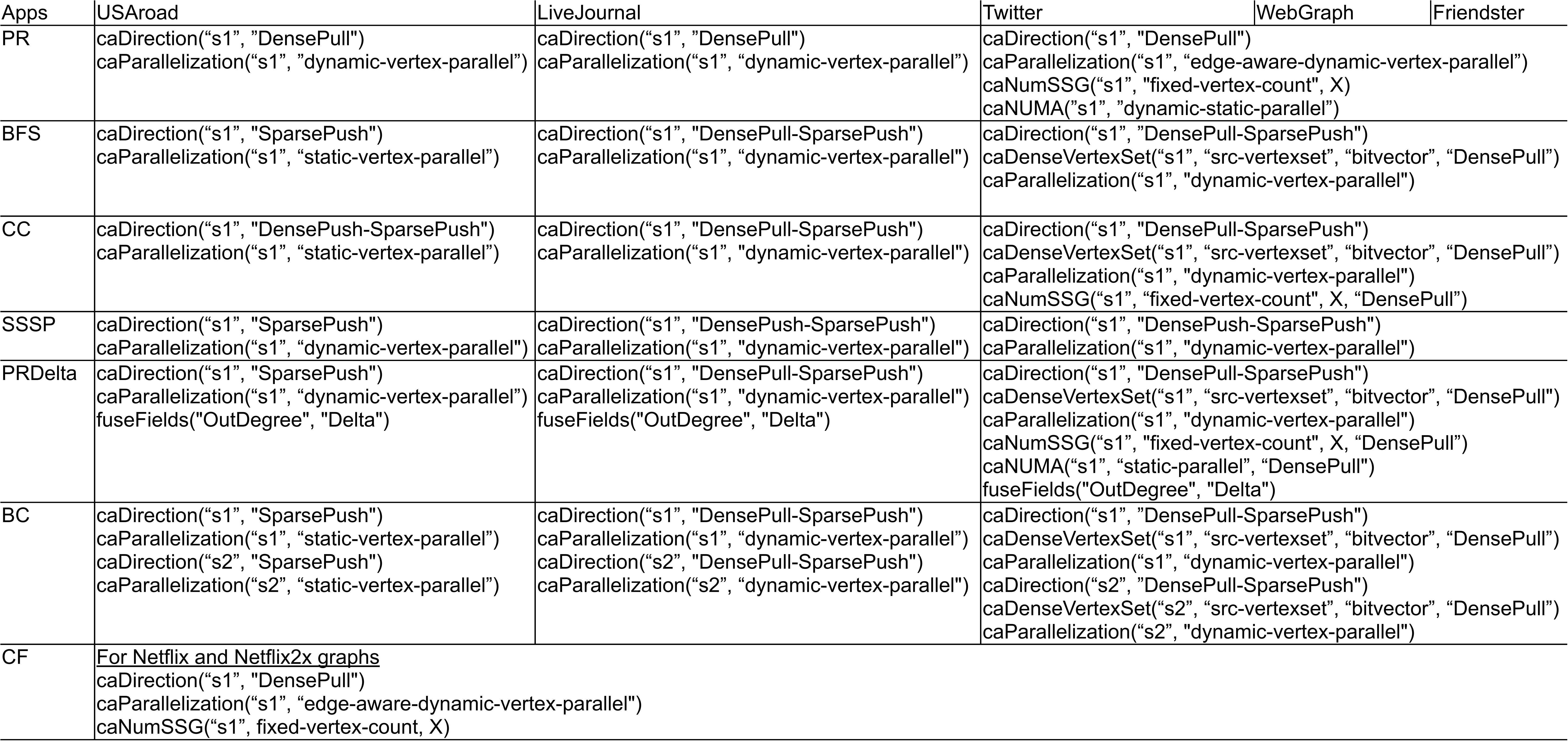}
\end{table}

\punt{
\julian{the font of this table is too small}
}

\begin{table}\scriptsize
  \tabcolsep 5pt
    \caption{Line counts of PR, BFS, CC, and SSSP for GraphIt, Ligra,
    GraphMat, Green-Marl, Galois, Gemini, and Grazelle. Only
    Green-Marl has fewer lines of code than GraphIt. GraphIt has an
    order of magnitude fewer lines of code than Grazelle (the second
    fastest framework on the majority of the algorithms we
    measured). For Galois, we only included the code for the specific
    algorithm that we used. Green-Marl has a built-in BFS.}
  \begin{tabular}{l|r|r|r|r|r|r|r}
    & GraphIt & Ligra & GraphMat & Green-Marl & Galois & Gemini & Grazelle \\ \hline
    PR & 34 & 74 & 140 & 20 & 114 & 127 & 388 \\
    BFS & 22 & 30 & 137 & 1 & 58 & 110 & 471  \\
    CC & 22 & 44 & 90 & 25 & 94 & 109 & 659 \\
    SSSP & 25 & 60 & 124 & 30 & 88 & 104 & \\ \hline
  \end{tabular}
  \label{table:LOC}
\end{table}

\begin{table}[!t]
  \center \tabcolsep 2pt \footnotesize
    \caption{LLC miss rate, QPI traffic, cycles with pending memory
    loads and cache misses, and parallel running time (seconds) of PR,
    CC, and PRDelta running on Twitter, and CF running on Netflix.}
  \label{table:perf-counters}
  \begin{tabular}{l|rrrr|rrrr|rr|rr}
    Algorithm & \multicolumn{4}{|c|}{ PR } & \multicolumn{4}{c|}{ CC } & \multicolumn{2}{c|}{ PRDelta } & \multicolumn{2}{c}{ CF } \\ \hline
    Metrics & GraphIt & Ligra & Gemini & Grazelle & GraphIt & Ligra & Gemini & Grazelle & GraphIt & Ligra & GraphIt & Ligra \\ \hline
    LLC miss rate (\%) & \textbf{24.59} & 60.97 & 45.09 & 56.68 & \textbf{10.27} & 48.92 & 43.46 & 56.24 & \textbf{32.96} & 71.16 & \textbf{2.82} & 37.86 \\
    QPI traffic (GB/s) & \textbf{7.26} & 34.83 & 8.00 & 20.50 & 19.81 & 27.63 & \textbf{6.20} & 18.96 & \textbf{8.50} & 33.46 & \textbf{5.68} & 19.64\\
    Cycle stalls (trillions) & \textbf{2.40} & 17.00 & 3.50 & 4.70 & \textbf{0.20} & 0.96 & 1.20 & 0.30 & \textbf{1.25} & 5.00 & \textbf{0.09} & 0.22 \\
    Runtime (s) & \textbf{8.71} & 49.00 & 10.98 & 15.70 & \textbf{0.89} & 2.78 & 3.85 & 1.73 & \textbf{4.72} & 9.19 & \textbf{1.29} & 5.35 \\ \hline
  \end{tabular}
\end{table}


\myparagraph{PR} GraphIt has the fastest PR on 4 out of the 5 graphs
and is up to 54\% faster than the next fastest framework because it
enables both cache and NUMA optimizations when necessary. 
Table~\ref{table:perf-counters} shows that on the Twitter
graph, \graphit has the lowest LLC misses, QPI traffic, and cycles
stalled compared to Gemini and Grazelle, which are the second and
third fastest. 
\graphit also reduces the line count by up to an order of magnitude compared to 
Grazelle and Gemini as shown in Table~\ref{table:LOC}. 
Grazelle uses the Vector-Sparse edge list to improve
vectorization, which works well on graphs with low-degree
vertices~\cite{Grossman2018}, outperforming \graphit by $23\%$ on
USAroad. 
\graphit does not yet have this optimization, but we plan to
include it in the future. Frameworks other than Gemini and Grazelle do
not optimize for cache or NUMA, resulting in much worse running times.

\myparagraph{BFS} \graphit has the fastest BFS on 4 out of the 5
graphs (up to 28\% faster than the next fastest) because of its
ability to generate code with different direction and bitvector
optimizations. On LiveJournal, Twitter, WebGraph, and Friendster,
\graphit adopts Ligra's direction optimization. On USAroad, \graphit always uses
SparsePush and omits the check for when to switch traversal direction,
reducing runtime overhead. In the pull direction traversals, 
\graphit uses bitvectors to represent the frontiers when
 boolean array representations do not fit in the last level cache, 
 whereas Ligra always
uses boolean arrays and Grazelle always uses bitvectors. GraphIt
outperforms Galois' BFS, even though Galois is highly-optimized for
road graphs. GraphMat and Green-Marl do not have the direction
optimization so it is much slower. Ligra is slightly faster than
\graphit on the smaller LiveJournal graph due to better memory
utilization, but is slower on larger graphs. 

\myparagraph{CC} GraphIt has the fastest CC on Twitter, WebGraph, and
Friendster because of the direction, bitvector, and cache
optimizations. Table~\ref{table:perf-counters} shows GraphIt's reduced
LLC miss rate and cycles stalled. Interestingly, Gemini has the lowest
QPI traffic, but is much slower than \graphit. With NUMA optimizations,
vertices in one socket fail to see the newly propagated labels from
vertices in another socket, resulting in slower convergence. Unlike
other NUMA-aware graph processing frameworks, \graphit can easily
enable or disable NUMA optimizations depending on the algorithm. We
choose the label propagation algorithm option on Galois and use the
FRONTIERS\_WITHOUT\_ASYNC option on Grazelle in order to compare the same
algorithm across frameworks.  Galois' CC is $35\%$ faster than \graphit
on USAroad because it uses a special asynchronous engine instead of a
frontier-based model. We also ran Galois's union-find CC
implementation but found it to be slower than \graphit on all graphs
except USAroad. Grazelle's CC using the Vector-Sparse format,
implemented with hundreds of lines of assembly code as shown in
Table~\ref{table:LOC}, is $43\%$ faster than \graphit on USAroad. The
best performing schedule that we found on USAroad without any
asynchronous mechanism is DensePush-SparsePush.

\myparagraph{CF} For CF, GraphIt is faster than Ligra and GraphMat (by
4--4.8$\times$) because the edge-aware-dynamic-vertex-parallel schedule
achieves good load balance on Netflix. Cache optimization further
improves GraphIt's performance and is especially beneficial on
Netflix2x.

\myparagraph{SSSP} \graphit has the fastest SSSP on 4 out of the 5
graphs because of its ability to enable or disable the direction
optimization and the bitvector representation of the frontier. We run
Galois with the Bellman-Ford algorithm so that the algorithms are the
same across systems. We also tried Galois's asynchronous SSSP but
found it to be faster than \graphit only on WebGraph.  Green-Marl's
SSSP on USAroad is 328 times slower than \graphit because it 
uses the DensePush configuration. On every round,
it must iterate through all vertices to check if they are
active. This is expensive on USAroad because for over 6000 rounds,
the active vertices count is less than $0.4\%$ of all the vertices. 

\myparagraph{PRDelta} GraphIt outperforms Ligra on all graphs by
2--4$\times$ due to better locality from using bitvectors as
frontiers, fusing the Delta and OutDegree arrays as shown in
\Fig~\ref{fig:code:pagerankdelta}, and applying both the cache and NUMA
optimizations in the pull direction. Table~\ref{table:perf-counters}
shows GraphIt's reduced LLC miss rate, QPI traffic, and cycles
stalled.

\updated{
\myparagraph{BC} \graphit achieves the fastest BC performance on the USAroad and
Friendster graphs and has comparable performance on the other graphs. 
\graphit is a bit slower than Gemini on Twitter and WebGraph because
it does not support bitvector as a layout option for 
vertex data vectors layouts. We plan to add this in the future.
} 

\subsection{Performance of Different Schedules}

\begin{figure}[t]
  \centering
  \includegraphics[width=0.8\columnwidth]{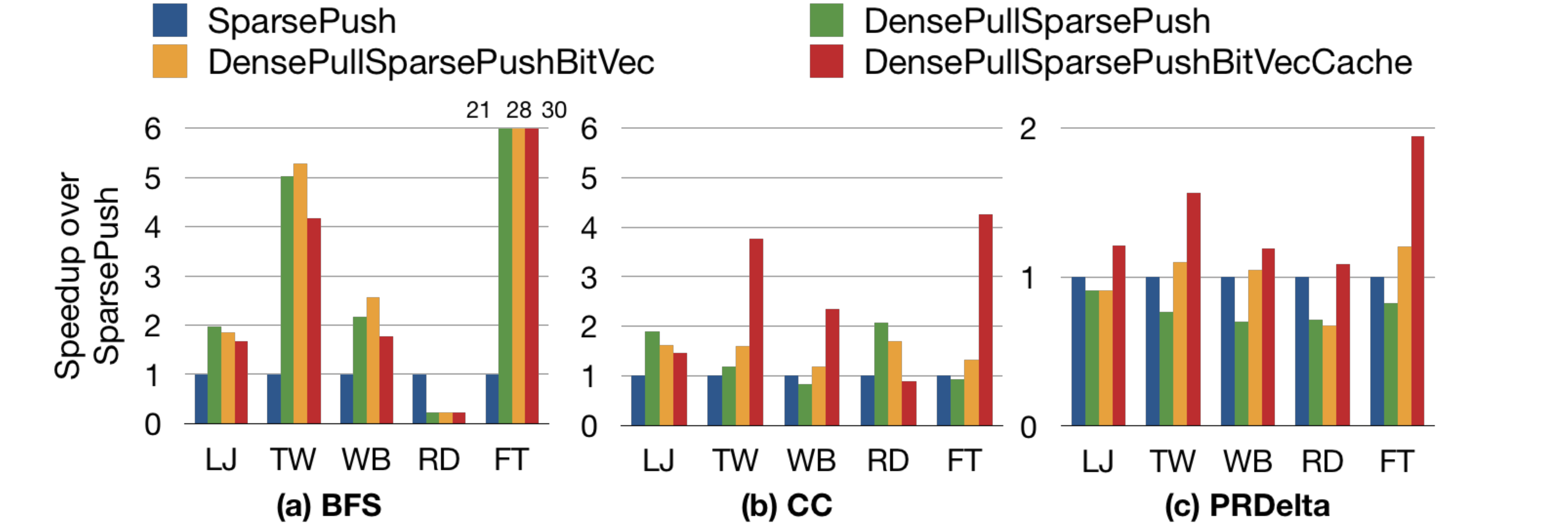}
  \caption{Performance of different schedules for BFS, CC, and
    PRDelta. SparsePush and DensePullSparsePush refer to the traversal
    directions. BitVec refers to the dense frontier data
    structure. Cache refers to the cache optimization. The
    descriptions of these schedules can be found in
    Section~\ref{subsec:tradeoff_optimizations}. The full scheduling
    commands are shown in Table~\ref{table:eval-schedules}.
  }
  \label{fig:perf_diff_schedules}
\end{figure}

\begin{figure}[t]
  \centering
  \includegraphics[width=0.7\columnwidth]{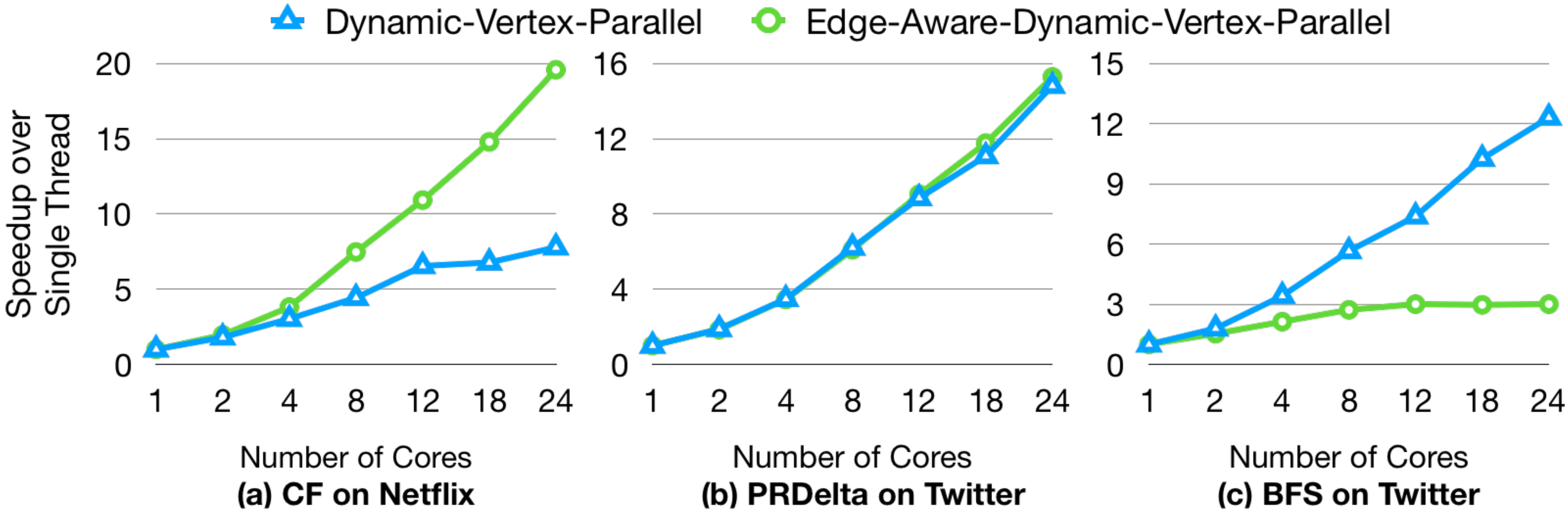}
  \caption{Scalability of CF, PRDelta, and BFS with different schedules. Hyper-threading is disabled. }
  \label{fig:scalability}
\end{figure}

\Fig~\ref{fig:perf_diff_schedules} demonstrates the impact
of traversal direction, data structures used for keeping track of
active vertices, and cache optimizations. For a given algorithm, there
is no single schedule that works well on all input graphs. For BFS,
DensePullSparsePush with cache optimizations reduces the number of
memory accesses on LiveJournal, Twitter, WebGraph, and Friendster,
achieving up to 30$\times$ speedup.  However, using only SparsePush
can reduce the runtime overhead on USAroad as described in
Section~\ref{subsec:compare_with_others}. For CC, the bitvector and
cache optimizations improve locality of memory accesses for Twitter,
WebGraph, and Friendster, but hurt the performance of LiveJournal and
USAroad due to lower \work. For PRDelta, SparsePush sometimes
outperforms DensePullSparsePush, but when the cache optimization is
applied to the pull direction, hybrid traversal is preferred.

\Fig~\ref{fig:scalability} shows that the parallelization scheme can
have a major impact on scalability, and again there is no single
scheme that works the best for all algorithms and inputs. For CF, the
amount of work per vertex is proportional to the number of edges
incident to that vertex. Consequently, the edge-aware-dynamic-vertex-parallel
scheme is 2.4$\times$ faster than the dynamic-vertex-parallel approach because
of better load balance.  For PRDelta, the number of active vertices
quickly decreases, and many of the edges do not need to be
traversed. As a result, the edge-aware-dynamic-vertex-parallel scheme has a
smaller impact on performance. The dynamic-vertex-parallel approach is a good
candidate for BFS because not all edges incident to a vertex are
traversed. Using the edge-aware-dynamic-vertex-parallel scheme for BFS ends up
hurting the overall load balance. We omit the edge-parallel approach
because it is consistently worse than edge-aware-dynamic-vertex-parallel due
to the extra synchronization overhead.

\updated{
\subsection{Autotuning Performance}
The autotuner found schedules that
 performed within 5\% of the hand-tuned schedules for 
all of the benchmarks in under 5000 seconds. 
For three benchmarks, the autotuner found schedules that outperformed 
hand-tuned schedules by up to 10\%. 
\punt{
The autotuner generally searched about 3\% of the total schedule 
space to find high-performance schedules. 
}

\punt{ \Fig~\ref{fig:perf_bfs} shows that the best schedule for a
  given algorithm can be different depending on the input. For social
  and web graphs, the hybrid schedule can be up to $19\times$ faster
  than a push only schedule. However, the push only schedule is
  $4\times$ faster for USAroad, due to reduction in runtime
  overhead. In contrast Ligra would always incur this overhead.

  This experiment shows that different classes of inputs require
  different schedules. Since \graphit separates the algorithm from the
  schedule, exploring different optimizations and applying the best
  among them becomes much easier.  No change on the algorithm is
  required; one only needs to modify the line of code that represents
  the schedule.  Performing the same optimization in another framework
  would have required the programmer to write a new implementation of
  the algorithm which is time consuming.  In \graphit, we provide the
  programmer with a powerful and flexible set of scheduling commands
  that they can use to explore the large space of possible
  optimizations without changing the algorithm.  }

  \subsection{Fusion of Multiple Graph Kernels}

\begin{figure}[t]
  \centering
  \includegraphics[width=0.8\columnwidth]{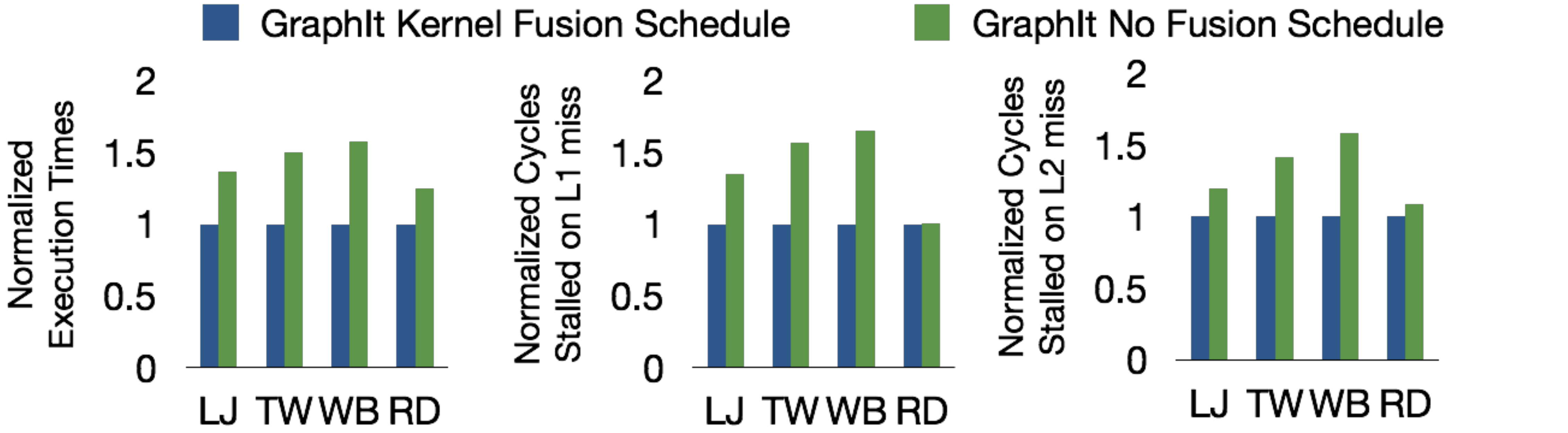}
  \caption{Normalized Execution Time, and L1 and L2 Cache Stall Cycles
    with Fusion of PageRank and Eigenvector Centrality}
  \label{fig:perf_fusion}
\end{figure}

\Fig~\ref{fig:perf_fusion} demonstrates the performance improvement of
kernel fusion with PageRank and Eigenvector Centrality. They have
similar memory access patterns. \graphit significantly improves the
spatial locality of the memory accesses by fusing together the two
kernels and the vectors they access (vertex data layout optimization). 
\Fig~\ref{fig:perf_fusion} shows 
significant reduction in cycles stalled on L1 data cache and L2 cache
misses, leading to the speedups.

\punt{ The ability to fuse loops and \apply functions and the ability
  to change data layouts is a unique advantage of the DSL approach
  that we take.  Performing such transformations in the general case
  and being able to compose them efficiently using a library is quite
  difficult and inefficient.  For example, applying a data layout
  transformation on a variable requires the modification of all the
  references to that variable which is difficult to be done
  efficiently without the use of compiler.  }
 }



\section{Related Work}
\label{sec:related}

 \begin{table}[!t]\scriptsize
\centering
\caption{Optimizations adopted by various frameworks
  (explained in Section \ref{sec:tradeoff}): WSVP (work-stealing
  vertex-parallel), WSEVP (work-stealing edge-aware vertex-parallel),
  SPVP (static-partitioned vertex-parallel with no work-stealing),
  EP (edge-parallel), BA (dense boolean array), BV (dense bitvector), AoS (Array
  of Structs), SoA (Struct of Arrays), SPS (SparsePush), DPS
  (DensePush), SP (SparsePull), DP (DensePull), SPS-DP (hybrid
  direction with SPS and DP depending on frontier size), DPS-SPS
  (hybrid with DPS and SPS), DPS-DP (hybrid with DPS and DP). }
\begin{tabular}[!t]{p{1.3cm}|p{1.3cm}|p{1.3cm}|p{1.3cm}|p{1.0cm}|p{1.25cm}|p{1.25cm}|p{1.42cm}|p{0.6cm}}
Frameworks & Traversal \newline Directions & Dense \newline Frontier \newline Data Layout & Parallelization & Vertex Data Layout & Cache \newline Opt. & NUMA \newline Opt. & Optimization Combinations Count & Integer Params Count\\
\hline\hline
GraphIt & SPS, DPS, SP, DP, SPS-DP, DPS-SPS & BA, BV & WSVP, WSEVP, SPVP, EP & AoS, SoA & Partitioned,\newline No Partition & Partitioned, Interleaved & \textbf{100+} & 3\\ \hline
Ligra & SPS-DP, DPS-SPS & BA & WSVP, EP & SoA & None & Interleaved & 4 & 1\\ \hline
Green-Marl & DPS, DP & BA & WSVP & SoA & None & Interleaved & 2 & 0\\ \hline
GraphMat & DPS, DP & BV & WSVP & AoS & None & Interleaved & 2 & 0\\ \hline
Galois  & SPS, DP, SPS-DP & BA & WSVP & AoS & None  & Interleaved & 3 & 1\\ \hline
Polymer & SPS-DP, DPS-SPS & BA & WSVP & SoA & None & Partitioned & 2 & 0\\ \hline
Gemini & SPS, DP, SPS-DP & BA,BV & WSVP & SoA & None & Partitioned & 6 & 1\\ \hline
GraphGrind & SPS-DP, DPS-SPS & BA & WSVP & SoA & None & Partitioned, Interleaved & 4 & 1 \\ \hline
Grazelle & DPS, DP, \newline DPS-DP & BV & EP & SoA & None & Partitioned & 3 & 1\\ \hline
\end{tabular}
\label{table:related-table}
\end{table}

\myparagraph{Shared-Memory Graph Processing Libraries and DSLs} Many
high-performance graph frameworks and DSLs, including \graphit,
optimize their performance for shared-memory systems. Many of these
frameworks support only a limited set of combinations of optimization
techniques as shown in Table~\ref{table:related-table} (these
optimizations are described in Section~\ref{sec:tradeoff}).
\emph{\graphit significantly expands the space of optimizations by
  composing large number of effective optimizations, supporting two
  orders of magnitude more optimization combinations than existing
  frameworks.}  \graphit achieves high performance by enabling
programmers to easily find the best combination of optimizations for
their specific algorithm and input graph.  \graphit also finds
previously unexplored combinations of optimizations to significantly
outperform state-of-the-art frameworks on many algorithms.

Many shared-memory graph systems, such as
Ligra~\cite{shun13ppopp-ligra}, Gunrock~\cite{Wang16gunrock},
GraphGrind~\cite{Sun2017}, Polymer~\cite{zhang15ppopp-numa-polymer},
Gemini~\cite{Zhu16gemni} and Grazelle~\cite{Grossman2018}, adopt the
frontier-based model. Galois~\cite{nguyen13sosp-galois} also has an
implementation of the model and a scheduler that makes it particularly
efficient for road graphs.  The frontier-based
model~\cite{shun13ppopp-ligra} operates efficiently on subsets of
vertices (frontiers) and their outgoing edges using the direction
optimization~\cite{Beamer-2012}.  Flat data-parallel operators are
used to apply functions to the frontier vertices and their neighbors
with parallel direction optimizations.  Existing frameworks only
support up to three of the many possible directions, with little
support for different parallelization schemes and frontier and vertex data layout optimizations.  \graphit
significantly expands the space of optimizations by enabling
combinations of data layout optimization, different direction choices,
and various parallelization schemes (Table~\ref{table:related-table}).
\graphit also makes programming easier by freeing the programmer from
specifying low-level implementation details, such as updated vertex
tracking and atomic synchronizations.

Many frameworks and techniques have been introduced to improve
locality with NUMA and cache optimizations. GraphGrind, Grazelle,
Gemini and Polymer all support NUMA optimizations. CSR
Segmenting~\cite{Yunming2017} and cache blocking~\cite{Nishtala2007,Beamer2017}
have been introduced to improve the cache performance of graph
applications through graph partitioning.  However, both techniques
have not been integrated in a general programming model or combined
with direction optimizations.  \graphit supports NUMA optimizations
and integrates a simplified variant of CSR segmenting to compose cache
optimizations with other optimizations.

Other shared-memory systems \cite{sundaram15vldb-graphmat,
  BigGraph2017} adopt the vertex-centric model 
  to exploit data parallelism across vertices.  Programmers
specify the logic that each (active) vertex executes iteratively.
  Frameworks~\cite{Pegasus,sdm12} use
sparse matrix-vector multiplication with semirings to
express graph algorithms. 
However, both programming
models cannot easily integrate direction optimization,
which requires different synchronization strategies for each vertex in
push and pull directions.

Green-Marl~\cite{Hong12asplos}, Socialite~\cite{socialite13ICDE}, Abelian~\cite{abelian2018}, and
EmptyHeaded~\cite{Aberger2016} are DSLs for shared-memory graph
processing. Green-Marl provides a BFS primitive, and so programs that
can be expressed with BFS invocations are relatively concise. However,
for other graph programs, the programmer needs to write the loops over
vertices and edges explicitly, making it hard to integrate direction
optimization due to the lower level nature of the language.  Socialite
and EmptyHeaded provide relational query languages to express graph
algorithms. The underlying data representation is in the form of
tables, and due to extensive research in join optimizations, these
systems perform especially well for graph algorithms that can be
expressed efficiently using joins (e.g., subgraph finding). However,
since these languages do not allow for explicit representation of
active vertex sets, their performance on graph traversal algorithms is
worse than the frontier-based
frameworks~\cite{Satish14Sigmod,Aberger2016}. These DSLs also do not
support composition of optimizations or expose extensive performance
tuning capabilities.

A number of graph processing frameworks have been developed for GPUs
(see~\cite{GPUSurvey2018} for a survey). We did not focus on GPUs in
this paper as the current GPU memory capacities do not allow us to
process very large graphs in-memory.

\myparagraph{Out-of-Core Graph Processing Frameworks} There has
been a significant amount of work dealing with graphs that cannot fit
in memory (e.g.,~\cite{kyrola12osdi-graphchi, zheng15fast-flashgraph,
  roy13sosp-xstream, Zhu15ATC-GridGraph, Maass2017, SanWoo2017,
  Vora2016, Wang15ATC, Graspan2017, BigGraph2017}), while \graphit
focuses on in-memory graph processing.  Some of the optimizations
 in out-of-core systems also focus on improving locality of
accesses, parallelism, and work-efficiency, but the tradeoff space for
these techniques is very different when optimizing for the disk/DRAM
boundary, instead of the DRAM/cache boundary.  The higher disk
access latency, lower memory bandwidth, and larger granularity of access
lead to very different techniques~\cite{SanWoo2017}.
When the graphs do fit in memory, out-of-core systems, 
such as X-Stream~\cite{roy13sosp-xstream}, have shown to be
much slower than shared-memory
frameworks~\cite{Yunming2017,Grossman2018}.

\myparagraph{Distributed Graph Processing Frameworks} Graph analytics
has also been studied extensively in distributed memory
systems~(e.g., \cite{low10uai-graphlab, Gonzalez2012, prabhakaran12atc-grace,
  Roy2015, Chen15PowerLyra, Sakr2017, Zhu16gemni, McCune2015,
  BigGraph2017, gluon2018}).  The tradeoff space is
also different for distributed graph processing systems due to the
larger network communication overhead and greater need for load
balance. Techniques used by \graphit, such as direction
optimization and locality enhancing graph partitioning can also be
applied in the distributed domain~\cite{Zhu16gemni}.  These systems,
when run on a single machine, generally cannot outperform
shared-memory frameworks~\cite{Satish14Sigmod}.

\myparagraph{Scheduling Languages}
\graphit introduces an
expressive scheduling language.  Examples of existing scheduling languages include
Halide~\cite{Ragan-Kelley:2013:HLC:2499370.2462176},
CHiLL~\cite{Chen08chill}, and HMPP~\cite{HMPP}.  These languages
mainly focus on loop nest optimization in applications that manipulate
dense arrays.
Unlike these scheduling languages, the \graphit scheduling language is
designed for graph applications.  It is the first
scheduling language designed to address the challenges of graph
applications, graph data structures, and graph optimizations.  It
allows the programmer to perform data layout transformations (which is not
possible in Halide), and allows full separation between the algorithm
and the schedule (which is not possible in HMPP).
Unlike CHiLL, which was designed mainly for the application of affine
transformations on loop nests, the GraphIt scheduling language
supports a large set of non-affine transformations, which are the main
type of optimizations in the context of graph applications.

\myparagraph{Program Synthesis} Program synthesis techniques have been
explored in the context of graph
algorithms~\cite{PrountzosMP12,PrountzosMP15}, which allow many
different implementations of an application to be generated. However,
little control is provided to compose together
different optimizations.  \graphit enables programmers to 
apply their knowledge to 
find profitable combinations of optimizations. \graphit also supports
a much wider range of optimizations.

\myparagraph{Physical Simulation DSLs} GraphIt is heavily
influenced by DSLs for physical simulations, including
Simit~\cite{fredSimit16} and Liszt~\cite{DeVito11Listz}.
 However, Simit and Liszt do not support efficient filtering on
vertices and edges, and do not have a scheduling language. 

\punt{
\graphit's
vector design is similar to Simit's system vector and Liszt's
field.}
\section{Conclusion}
\label{sec:conclusion}

\updated{
We have described \graphit, a novel DSL for graph processing that
generates fast implementations for algorithms
with different performance characteristics running on graphs
with varying sizes and structures.
\graphit separates algorithm specifications from performance optimizations. 
The algorithm language simplifies expressing the algorithms.
We formulate graph optimizations as tradeoffs among 
locality, parallelism, and \work. 
The scheduling language enables programmers
to easily search through the complicated
tradeoff space.
We introduce the
\textit{graph iteration space} to model,
compose, and ensure the validity of the edge traversal optimizations. 
The separation of algorithm and schedule, and the correctness
guarantee of edge traversal optimizations 
 enabled us to build an autotuner on top 
of \graphit. 
Our experiments show that
 \graphit is up to 4.8$\times$ faster than state-of-the-art graph frameworks.
Future work includes extending the compiler to support 
more optimizations and hardware platforms (e.g., GPU and distributed-memory).
} 

\section*{Acknowledgments}

We thank Tyler Denniston, Vladimir Kiriansky, Jure Leskovec, Michael
W. Mahoney, and the reviewers for their helpful feedback and
suggestions.  This research was supported by Toyota Research
Institute, DoE Exascale award \#DE-SC0008923, DARPA SDH Award
\#HR0011-18-3-0007, DARPA D3M Award \#FA8750-17-2-0126, Applications
Driving Architectures (ADA) Research Center, a JUMP Center
co-sponsored by SRC and DARPA, and DOE Early Career Award
\#DE-SC0018947.

\bibliography{graph}

\punt{ 
\newpage
\appendix
\section{Full code for Ligra's PageRankDelta}

  \begin{lstlisting} [language=c++,escapechar=|]
template <class vertex>
struct PR_F {
  vertex* V;
  double* Delta, *nghSum;
  PR_F(vertex* _V, double* _Delta, double* _nghSum) : 
    V(_V), Delta(_Delta), nghSum(_nghSum) {}
  inline bool update(uintE s, uintE d){
    double oldVal = nghSum[d];
    nghSum[d] += Delta[s]/V[s].getOutDegree();
    return oldVal == 0;
  }
  inline bool updateAtomic (uintE s, uintE d) {
    volatile double oldV, newV; 
    do { //basically a fetch-and-add
      oldV = nghSum[d]; newV = oldV + Delta[s]/V[s].getOutDegree();
    } while(!CAS(&nghSum[d],oldV,newV));
    return oldV == 0.0;
  }
  inline bool cond (uintE d) { return cond_true(d); }};

struct PR_Vertex_F_FirstRound {
  double damping, addedConstant, one_over_n, epsilon2;
  double* p, *Delta, *nghSum;
  PR_Vertex_F_FirstRound(double* _p, double* _Delta, double* _nghSum, double _damping,
    double _one_over_n,double _epsilon2) :
    p(_p), damping(_damping), Delta(_Delta), nghSum(_nghSum), one_over_n(_one_over_n),
    addedConstant((1-_damping) * _one_over_n), epsilon2(_epsilon2) {}
  inline bool operator () (uintE i) {
    Delta[i] = damping * nghSum[i] + addedConstant;
    p[i] += Delta[i];
    Delta[i] -= one_over_n; //subtract off delta from initialization
    return (fabs(Delta[i]) > epsilon2 * p[i]);
  }
};

struct PR_Vertex_F {
  double damping, epsilon2;
  double* p, *Delta, *nghSum;
  PR_Vertex_F(double* _p, double* _Delta, double* _nghSum, double _damping, double _epsilon2) :
    p(_p), damping(_damping), Delta(_Delta), nghSum(_nghSum), epsilon2(_epsilon2) {}
  inline bool operator () (uintE i) {
    Delta[i] = nghSum[i] * damping;
    if (fabs(Delta[i]) > epsilon2 * p[i]) { p[i] += Delta[i]; return 1;}
    else return 0;
  }
};

struct PR_Vertex_Reset {
  double* nghSum;
  PR_Vertex_Reset(double* _nghSum) :
    nghSum(_nghSum) {}
  inline bool operator () (uintE i) {
    nghSum[i] = 0.0;
    return 1;
  }
};

template <class vertex>
void Compute(graph<vertex>& GA, commandLine P) {
  long maxIters = P.getOptionLongValue("-maxiters",100);
  const long n = GA.n;
  const double damping = 0.85;
  const double epsilon = 0.0000001;
  const double epsilon2 = 0.01;
  double one_over_n = 1/(double)n;
  double* p = newA(double,n), *Delta = newA(double,n), *nghSum = newA(double,n);
  bool* frontier = newA(bool,n);
  parallel_for(long i=0;i<n;i++) {
    p[i] = 0.0;//one_over_n;
    Delta[i] = one_over_n; //initial delta propagation from each vertex
    nghSum[i] = 0.0;
    frontier[i] = 1;
  }

  vertexSubset Frontier(n,n,frontier);
  bool* all = newA(bool,n);
  {parallel_for(long i=0;i<n;i++) all[i] = 1;}
  vertexSubset All(n,n,all); //all vertices

  long round = 0;
  while(round++ < maxIters) {
    edgeMap(GA,Frontier,PR_F<vertex>(GA.V,Delta,nghSum),GA.m/20, no_output $\mid$ dense_forward);
    vertexSubset active 
      = (round == 1) ? 
      vertexFilter(All,PR_Vertex_F_FirstRound(p,Delta,nghSum,damping,one_over_n,epsilon2)) :
      vertexFilter(All,PR_Vertex_F(p,Delta,nghSum,damping,epsilon2));
    //compute L1-norm (use nghSum as temp array)
    {parallel_for(long i=0;i<n;i++) {
      nghSum[i] = fabs(Delta[i]); }}
    double L1_norm = sequence::plusReduce(nghSum,n);
    if(L1_norm < epsilon) break;
    //reset
    vertexMap(All,PR_Vertex_Reset(nghSum));
    Frontier.del();
    Frontier = active;
  }
  Frontier.del(); free(p); free(Delta); free(nghSum); All.del();
}
\end{lstlisting}

}
\end{document}